\documentclass[11pt]{article}
\usepackage{graphicx}
\usepackage[authoryear,round]{natbib}

\usepackage{amsmath, amssymb, amsfonts, amsthm, mathtools}
\usepackage{multirow, multicol}
\usepackage{enumerate, array, enumitem}
\usepackage{epsfig, graphics}
\usepackage{float, lscape}
\usepackage{color, xcolor}
\usepackage{verbatim}
\usepackage{lineno}
\usepackage[pdfstartview=XYZ,bookmarks=true,colorlinks=true,
  linkcolor=blue,urlcolor=blue,citecolor=blue,linktocpage=true,hyperindex=true]{hyperref}
\usepackage{xr-hyper}
\usepackage[nameinlink]{cleveref}
\usepackage{caption}
\usepackage{url}
\usepackage{booktabs}
\usepackage{pdfpages}
\newtheorem{proposition}{Proposition}[section]

\theoremstyle{remark}

\makeatletter

\newcommand{\beq}{\begin{equation}}
\newcommand{\eeq}{\end{equation}}

\makeatletter

\newcommand{\pushright}[1]{\ifmeasuring@#1\else\omit\hfill$\displaystyle#1$\fi\ignorespaces}
\newcommand{\pushleft}[1]{\ifmeasuring@#1\else\omit$\displaystyle#1$\hfill\fi\ignorespaces}
\makeatother

\renewcommand{\baselinestretch}{1.2}

\begin{document}

	\title{Automated selection of $r$ in $r$-largest order statistics models:
		finite-sample performance and return-level inference}
	
	\author{%
		Yire Shin$^{1}$,~Jihong Park$^{2,*}$,~Jeong-Soo Park$^{1}$ \\[2ex]
		\small \it $^1$ Department of Statistics, Chonnam National University, Gwangju 61186, Korea \\
		\small \it $^2$ Department of Mathematics and Statistics, \\ \small \it Chonnam National University, Gwangju 61186, Korea \\
		\small \it *Corresponding author: jihong8090@gmail.com   
	}
		\maketitle

\begin{abstract}
	Selecting the number $r$ of upper order statistics is a key practical issue in fitting $r$-largest order statistics models. The entropy difference test for selecting $r$ relies on a large-sample approximation. The score test can be used for small samples but is computationally demanding. To obtain a fast method that is also applicable to small samples, we propose a method based on the conditional cumulative distribution function (CCDF), which applies a sequence of Cram\'er--von Mises goodness-of-fit tests. The tests are calibrated by a parametric bootstrap in which each refit is replaced by a single Newton step, so the computation remains fast. We also revisit the spacings test and show that the two test statistics become nearly identical as $r$ increases. Monte Carlo simulations show that the CCDF test has accurate size and good power in small and moderate samples. Further simulations show that a few additional order statistics can reduce the error of return-level estimates compared with those obtained from the block maxima only. The method is illustrated using rainfall data from Sancheong, Korea.
\end{abstract}
	
\vspace{5mm} \noindent {\bf Keywords}: {Cram{\'e}r--von Mises test; one-step calibration; $r$-largest order statistics; parametric bootstrap; return level;
		sequential testing} 

\section{Introduction}

Accurate inference for extreme upper quantiles is important in risk
assessment and planning for rare events. Two standard approaches are
the block maxima (BM) method, modeled by the generalized extreme-value (GEV)
distribution, and the peaks-over-threshold method, modeled by the generalized
Pareto distribution 
\citep{coles2001introduction,
	beirlant2006book}.
The BM approach is simple and well established but uses only one
observation from each block, whereas the peaks-over-threshold method uses more
upper-tail observations at the cost of selecting an appropriate
threshold. Their relative performance depends on the data structure
and inferential objective
\citep{Tabari_dilemma2021,BucherZhou2021}.

The $r$-largest order statistics (rLOS) approach provides an
intermediate framework by using the largest $r$ observations from each
block. The model was developed for the Gumbel case by
\citet{smith1986extreme}, building on \citet{Weissman1978}, and was
extended to an arbitrary shape parameter by \citet{tawn1988extreme}.
The resulting $r$-largest GEV distribution ($r$GEV) retains the three parameters of the
ordinary GEV model and has been used in a range of environmental and
engineering applications
\citep{soares2004application,unnikrishnan2004analysis,
	an2007r,wang2008downscaling,hamdi2014extreme,
	muhammed2017variations,e2020change}.

A key practical issue is the choice of $r$. Small values of $r$
use less information and typically result in larger standard errors,
whereas increasing $r$ can improve estimation efficiency but may
introduce bias when the $r$GEV approximation becomes inadequate
\citep{coles2001introduction}. The selection of
$r$ thus involves a bias--variance trade-off
\citep{bader2017automated}.

Several methods have been proposed for selecting $r$.
\citet{smith1986extreme} and \citet{tawn1988extreme} used probability
plots based on marginal distributions, while \citet{tawn1988extreme}
also suggested a diagnostic based on the exponential-spacing property
of upper order statistics \citep{Weissman1978}. Other proposals include
the method of \citet{wang1995selection}, which requires information
about the domain of attraction, and the robust procedure of
\citet{dupuis1997extreme}, which involves Monte Carlo integration.

More recently, \citet{bader2017automated} proposed two tests within a
sequential hypothesis testing framework: the score test and the entropy
difference (ED) test. For the score test, the usual asymptotic $\chi^2$
distribution is not applicable, so a parametric bootstrap is used to
calibrate the test statistic. The ED test is based on the difference in
estimated entropy between the $r$GEV and $(r-1)$GEV models. In the
simulation study of \citet{bader2017automated}, the ED test showed
better power than the score test, and it has subsequently been used in
several applications
\citep{silva2022dynamic,Busaba2025CSAM,ShinPark2024Serra}. However, the ED test relies on a large-sample
approximation and is recommended for relatively large samples. For
small samples, the score test with a parametric bootstrap is
recommended, but its computational cost can be substantial. This
motivates a computationally efficient method for selecting $r$ that
also performs well in small samples.

We propose a method based on a sequence of goodness-of-fit tests using
the conditional cumulative distribution function (CCDF). Under the
$r$GEV model, the conditional distribution of the $r$th order statistic
given the preceding $r-1$ order statistics is a GEV distribution
right-truncated at the $(r-1)$th order statistic
\citep{bader2017automated}. The corresponding CCDF values therefore
follow a uniform $(0,1)$ distribution. We test this property
sequentially as $r$ increases from 2 to a prespecified maximum order
$R$, using the Cram\'er--von Mises (CvM) goodness-of-fit test.

The conditional probability-integral transform, also known as the
Rosenblatt transform, is classical and has long been used in
goodness-of-fit assessment, particularly for copula models
\citep{DobricSchmid2007,BergBakken2007,hofert2014graphical}.
Here we apply it to each newly added order statistic, which yields a
sequential test for selecting $r$.

We also revisit the spacings test, which is based on the
exponential-spacing property suggested by \citet{tawn1988extreme}.
We compare the CCDF test with the spacings test, the score test, and
the ED test through Monte Carlo simulations. In addition, we show that
the CCDF and spacings statistics are closely related: their
correlation tends to one as $r$ increases. This relationship explains
the similar $r$-selection behavior of the two tests.

A second contribution concerns the computation of $p$-values. In the CCDF
test, the parameters are estimated from the same sample that is used to
compute the CCDF values. As a result, the usual CvM $p$-values are too
large, and the test is conservative. The $p$-values therefore need to be
calibrated. A standard approach is the parametric bootstrap
\citep{GenestRemillard2008}. However, the
parametric bootstrap requires the MLE to be recomputed for every
bootstrap sample, which is time-consuming. Instead of computing the
MLE, we update the original estimate by a single Newton step for each
bootstrap sample \citep{LeCam1956,Andrews2002}. In our Monte Carlo experiments, this one-step bootstrap
gave nearly the same $p$-values as the full bootstrap, while the
parameter estimation step was about 65--146 times faster.

Finally, we study how the choice of $r$ affects the estimation of
return levels. A goodness-of-fit test selects the largest $r$ for which
the $r$GEV model fits the data. This value is not necessarily the best
choice for estimating a particular return level. Our simulations show
that using a few more order statistics than the largest correctly
specified order can reduce the root mean squared error of the estimated
100-year return level. The additional order statistics reduce the
variance of the estimate more than they increase its bias.
The proposed methods are illustrated using the largest daily rainfall
amounts in each year at Sancheong, Korea. As a further illustration,
the CCDF test is applied to the four-parameter kappa $r$-largest model
of \citet{shin2023modeling}, which shows that the method can be used
beyond the $r$GEV model.

The remainder of the paper is organized as follows.
Section~2 introduces the $r$GEV model and parameter estimation.
Section~3 reviews existing selection methods and develops the spacings
and CCDF procedures. Section~4 presents their theoretical relationship
and develops the one-step parametric-bootstrap approximation.
Section~5 reports the Monte Carlo experiments. Section~6 presents the
Sancheong rainfall application, and Sections~7 and 8 contain the
discussion and conclusions. Additional results and computational
details are provided in the Supplementary Information.

\section{Model and parameter estimation}

Let
\[
\underline{X}^{\,r}
=
(X^{(1)},\ldots,X^{(r)})
\]
denote the $r$ largest order statistics (rLOS) within a block, ordered as
\[
X^{(1)}\ge X^{(2)}\ge\cdots\ge X^{(r)}.
\]
The joint probability density function of the $r$GEV distribution is
\citep{tawn1988extreme,coles2001introduction}
\begin{linenomath*}
	\begin{equation}
	\label{rgevd}
	f_r(x^{(1)},\ldots,x^{(r)};\theta)
	=
	\sigma^{-r}F(x^{(r)})
	\prod_{s=1}^{r}
	w(x^{(s)})^{1/k-1},
	\end{equation}
\end{linenomath*}
where $\theta=(\mu,\sigma,k)$,
\begin{equation}
\label{wx}
\begin{aligned}
F(x)
&=
\exp\left\{-w(x)^{1/k}\right\},\\
w(x)
&=
1-k\frac{x-\mu}{\sigma},
\end{aligned}
\end{equation}
with $x^{(1)}\ge\cdots\ge x^{(r)}$, $\sigma>0$, and $w(x^{(s)})>0$ for $s=1,\ldots,r$.
Here, $\mu$, $\sigma$, and $k$ are the location, scale, and shape
parameters, respectively. The case $k=0$ is understood by continuity
and corresponds to the $r$-largest Gumbel model. Our sign convention for
$k$ is opposite to the commonly used GEV shape parameter $\xi$; that is,
$k=-\xi$ \citep{coles2001introduction,bader2017automated}.

The density in Eq.~(\ref{rgevd}) is a limiting approximation obtained as
the block size $B\rightarrow\infty$. The approximation is appropriate when $r$ is
small relative to $B$, because the rate of convergence to the limiting
joint distribution in Eq.~(\ref{rgevd}) decreases sharply as $r$
increases \citep{smith1986extreme}. Throughout, the rLOS vectors from different
blocks are assumed to be independent and identically distributed.

\subsection{Maximum likelihood estimation}

Let
\[
\underline{x}^{\,r}_i
=
(x_i^{(1)},\ldots,x_i^{(r)}),
\qquad i=1,\ldots,n,
\]
denote the observed rLOS vectors from $n$ independent blocks, for a
fixed $r$.
For $k\ne0$, the log-likelihood is
\begin{equation}
\label{gevnllh}
\begin{aligned}
\ell_n^{(r)}(\theta)
=
\sum_{i=1}^{n}\ell_i^{(r)}(\theta)
=
-\sum_{i=1}^{n}
\bigg[
&r\log\sigma
+
\{w(x_i^{(r)})\}^{1/k}\\
&+
\left(1-\frac{1}{k}\right)
\sum_{s=1}^{r}
\log\{w(x_i^{(s)})\}
\bigg],
\end{aligned}
\end{equation}
subject to
\[
w(x_i^{(s)})>0,
\qquad
i=1,\ldots,n,\quad s=1,\ldots,r.
\]
The maximum likelihood estimator (MLE),
$\widehat{\theta}_n^{(r)}$, is obtained by maximizing
Eq.~(\ref{gevnllh}). When $r$ is fixed, we suppress the superscript
$(r)$ where no ambiguity arises. 
For numerical optimization, positivity of the scale parameter is
enforced through the reparameterization $\eta=\log\sigma$.
Candidate parameter values violating the support condition
$w(x_i^{(s)})>0$ for any observation are treated as inadmissible
and excluded from the optimization.
The MLE for rLOS models can be computed using the R packages
\texttt{ismev} \citep{ismev2025}, \texttt{eva} \citep{bader2020package},
and \texttt{evmr} \citep{ShinParkRpackageevmr}.

Several alternatives to ordinary maximum likelihood have recently been
proposed for $r$GEV models, including a hybrid 
estimator \citep{Busababodhin2026}. 
For the $r$GEV analyses in this study, we use maximum likelihood
estimation.

\section{Methods for selecting $r$}
\label{sec:selection}

To select $r$, we sequentially assess the goodness-of-fit of the
$r$GEV model as $r$ increases, following \citet{bader2017automated}.
For each candidate order $r \ge 2$, we consider
\begin{equation}
\label{hypotheses}
\begin{aligned}
H_0^{(r)}:\quad&
\text{the $r$GEV model provides an adequate fit to the $r$LOS data},\\
H_1^{(r)}:\quad&
\text{the $(r-1)$GEV model provides an adequate fit to the
	$(r-1)$LOS data,}\\
&\text{but the $r$GEV model does not adequately fit the $r$LOS data}.
\end{aligned}
\end{equation}

The hypotheses are tested sequentially as $r$ increases from 2 to a
prespecified maximum order $R$. The procedure stops at the first
rejection of $H_0^{(r)}$, and the preceding order $r-1$ is selected.
We select $r-1$ because the rate of convergence to the $r$GEV
distribution decreases sharply as $r$ increases
\citep{smith1986extreme}, so that a model that fails at order $r$ is
not expected to fit at larger orders. The first-rejection rule is a
practical selection rule; it does not guarantee that the model is
inadequate at every order larger than $r$. If no rejection occurs up to
$R$, we set $\widehat r=R$. For $r=1$, a goodness-of-fit test of the
GEV model for the block maxima can also be reported as a preliminary
check.

\subsection{Existing score and entropy-difference tests}

\paragraph{Score test.}
\cite{bader2017automated} proposed a goodness-of-fit test based on the
score statistic
\begin{equation}
\label{score_stat}
V_n
=
\frac{1}{n}
S^\top(\widehat{\theta}_n)\,
I^{-1}\,
S(\widehat{\theta}_n),
\end{equation}
where $S(\theta)=\sum_{i=1}^n\partial\ell_i^{(r)}(\theta)/\partial\theta$
is the score function of the sample, $I=I(\widehat{\theta}_n)$ is the
Fisher information matrix per block \citep{tawn1988extreme}, and both are
evaluated at the MLE $\widehat{\theta}_n$. As noted by
\citet{bader2017automated}, the support of the $r$GEV distribution
depends on the parameters unless $k=0$, so the classical regularity
conditions for the score test \citep{CasellaBerger2nded} do not hold,
and $V_n$ need not follow a $\chi^2$ distribution with three degrees of
freedom asymptotically. Nevertheless, $V_n$ can be used to assess
goodness-of-fit, because the score function has expectation zero under
$H_0^{(r)}$. A large value of $V_n$ relative to its sampling
distribution indicates lack of fit and provides evidence against
$H_0^{(r)}$. To approximate the sampling distribution of $V_n$,
\citet{bader2017automated} considered two approaches: the parametric
bootstrap and the multiplier bootstrap
\citep[e.g.][]{kojadinovic2012goodness}. The multiplier bootstrap is
computationally faster than the parametric bootstrap, but it is
applicable only to large samples.

\paragraph{Entropy-difference test.}
The entropy-difference (ED) test of \cite{bader2017automated} is based
on the change in log-likelihood contribution when the $r$th order
statistic is added to the first $r-1$ order statistics. Let $\ell_i^{(r)}$ denote the log-likelihood
contribution of block $i$ under the $r$GEV model, and define, for
$r\ge2$,
\[
D_{ir}(\theta)
=
\ell_i^{(r)}(\theta)-\ell_i^{(r-1)}(\theta).
\]
The ED statistic is
\begin{equation}
\label{test_stat}
T_{rn}(\theta)
=
\frac{\sqrt{n}\{\bar{D}_{rn}-\eta_r\}}
{S_{D_{rn}}},
\end{equation}
where
\[
\bar{D}_{rn}
=
\frac{1}{n}\sum_{i=1}^nD_{ir},
\qquad
S_{D_{rn}}^2
=
\frac{1}{n-1}
\sum_{i=1}^n
(D_{ir}-\bar{D}_{rn})^2,
\]
and
\begin{equation}
\label{etar}
\eta_r
=
E(D_{1r})
=
-\log\sigma-1+(1-k)\psi(r),
\end{equation}
with $\psi(z)=d\log\Gamma(z)/dz$ denoting the digamma function.
Under $H_0^{(r)}$,
\[
T_{rn}(\theta)\xrightarrow{d}N(0,1),
\qquad n\rightarrow\infty.
\]
In practice, $\theta$ is replaced by the MLE $\widehat{\theta}$ under
$H_0^{(r)}$, and the two-sided $p$-value is computed from
$T_{rn}(\widehat{\theta})$.

The ED test is computationally inexpensive, but its calibration relies
on a large-sample normal approximation. \citet{bader2017automated}
reported that this approximation is satisfactory for samples as small
as $n=50$ when $r<5$ and $k\le0$, and that $n=100$ seems sufficient for
$r$ up to 10. For samples smaller than 50, they recommended the score
test with a parametric bootstrap. This test holds its level well for
samples as small as $n=25$, but it is computationally demanding,
because the MLE must be recomputed for every bootstrap sample.

The R package \texttt{eva} \citep{bader2020package} implements the ED
and score tests, together with the ForwardStop and StrongStop
procedures for ordered hypothesis testing \citep{gsell2016sequential}
through its \texttt{pSeqStop} function. Following the sequential
testing strategy of \citet{bader2017automated}, we first use the
unadjusted $p$-values and stop at the first rejection. If these do not
lead to a selection of $r$, we apply ForwardStop and then StrongStop.
Because the sequential $p$-values are dependent and do not necessarily
satisfy the assumptions of these stopping rules, the ForwardStop and
StrongStop results should be regarded as supplementary. The
unadjusted first-rejection rule is a screening rule that identifies the first order at
which the sequential path shows evidence of lack of fit; it does not
provide familywise error control.

\subsection{Spacings test}
\label{sec:spacing}

A classical exponential-spacing property provides another way to assess
the adequacy of the $r$GEV model. Define
\begin{equation}
\label{rGum}
Y_r
=
-\frac{1}{k}
\log\left\{
1-k\frac{X^{(r)}-\mu}{\sigma}
\right\},
\qquad r=1,\ldots,R.
\end{equation}
Then $(Y_1,\ldots,Y_R)$ follows the $R$-largest standard
Gumbel distribution; a short derivation is given in
Appendix~\ref{app:gumbel_transform}.

For $r\ge2$, define
\begin{equation}
\label{spacing_exp}
S_r(\theta)
=
(r-1)(Y_{r-1}-Y_r)
=
\frac{r-1}{k}
\log\left\{
\frac{\sigma-k(X^{(r)}-\mu)}
{\sigma-k(X^{(r-1)}-\mu)}
\right\}.
\end{equation}
Under the $r$GEV model, the spacings $Y_{r-1}-Y_r$ are independent
exponential random variables with mean $1/(r-1)$
\citep[Theorem~3]{Weissman1978}; see also
\citet[p.~202]{embrechts2013modelling}. Hence
$S_r(\theta)\sim\mathrm{Exp}(1)$.

For known $\theta$, the transformation
\begin{equation}
\label{spacing_unif}
V_r(\theta)
=
\exp\{-S_r(\theta)\}
\end{equation}
is uniform on $(0,1)$ under the $r$GEV model. The goodness-of-fit of
the $r$GEV model can therefore be assessed by applying the CvM test to
the values $V_{i,r}(\theta)$, $i=1,\ldots,n$. In practice,
$\theta$ is replaced by its estimate under $H_0^{(r)}$.
Because parameter estimation affects the finite-sample null
distribution, the resulting CvM statistic is calibrated as described
in Section~\ref{sec:cvm_selection}.

The exponential-spacing diagnostic was suggested by
\cite{tawn1988extreme}, but was not developed there as an automated
procedure for selecting $r$. Here we embed it in the same sequential
testing framework used for the other methods.

\subsection{Conditional CDF test: Proposed method}
\label{sec:ccdf}

For $2\le r\le R$, the conditional CDF of $X^{(r)}$ given the
preceding order statistics is
\begin{equation}
\label{ccdf}
G_\theta(x^{(r)}\mid\underline{x}^{\,r-1})
=
\frac{F_\theta(x^{(r)})}
{F_\theta(x^{(r-1)})},
\qquad
x^{(r)}\le x^{(r-1)},
\end{equation}
where $F_\theta$ is the GEV distribution function given in
Eq.~(\ref{wx}). Equivalently, this is the CDF of a GEV distribution
right-truncated at $x^{(r-1)}$ \citep{johnson1995continuous}.
Equation~(\ref{ccdf}) also exhibits the Markov property; the
conditional distribution of $X^{(r)}$ given the preceding order
statistics depends only on $X^{(r-1)}$.

Define
\begin{equation}
\label{Ur}
U_r(\theta)
=
G_\theta
\left(
X^{(r)}\mid X^{(r-1)}
\right),
\qquad r=2,\ldots,R,
\end{equation}
and let
\[
U_1(\theta)=F_\theta(X^{(1)}).
\]
At the true parameter value,
\[
U_1(\theta),\ldots,U_r(\theta)
\]
are independent uniform $(0,1)$ random variables. This is the
Rosenblatt transform \citep{rosenblatt1952remarks} applied to the
successive order statistics.

Rosenblatt transforms have been used for goodness-of-fit
assessment in copula models
\citep{DobricSchmid2007,BergBakken2007,hofert2014graphical}. In recent
independent work, \citet{heinonen7451430dependence} use a related
construction, based on the parameter-free copula of the largest order
statistics, to detect serial dependence between extremes. Our approach
differs from these studies. It does not use a copula representation,
but follows the successive structure of the order statistics, in which
the conditional distribution of $X^{(r)}$ depends only on $X^{(r-1)}$.
The novelty is therefore not the Rosenblatt transform itself, but its
use in a sequential test that decides whether each additional order
statistic should be retained, and thus selects $r$ automatically.

For a sample of $n$ independent blocks, we compute
\[
U_{i,r}(\widehat{\theta}^{(r)}),
\qquad i=1,\ldots,n,
\]
where $\widehat{\theta}^{(r)}$ is estimated under $H_0^{(r)}$, and test
whether these values are uniform on $(0,1)$. Because $\theta$ is
estimated from the same data that are used to compute the CCDF values,
the $p$-values are calibrated by the parametric bootstrap described in
Section~\ref{sec:cvm_selection}. For $r=1$, the CCDF value reduces to
the GEV distribution function evaluated at the block maximum, so the
test becomes an ordinary goodness-of-fit test of the GEV model.

\subsection{Cram{\'e}r--von Mises calibration and sequential selection}
\label{sec:cvm_selection}

Both the spacings and CCDF procedures can be expressed as tests of
uniformity. For the CCDF method, the transformed observations are
$U_{i,r}(\widehat{\theta}^{(r)})$, while for the spacings method they are
$V_{i,r}(\widehat{\theta}^{(r)})$ from Eq.~(\ref{spacing_unif}).
For a fixed $r$, let $F_{n,r}$ denote the empirical distribution
function of the $n$ transformed values $U_{1,r},\ldots,U_{n,r}$ for the
CCDF test, or of $V_{1,r},\ldots,V_{n,r}$ for the spacings test, computed
with $\widehat{\theta}^{(r)}$.

We use the Cram\'er--von Mises (CvM) test for uniformity. Other
goodness-of-fit tests, such as the Anderson--Darling or
Kolmogorov--Smirnov test, could also be used; in our small-scale
experiments (not reported here), they performed similarly.
\citet{bader2018automated} used the Anderson--Darling test for
automated threshold selection.
The Cram\'er--von Mises statistic is
\begin{equation}
\label{cvm_stat}
W_{n,r}^{2}
=
n\int_0^1
\left\{
F_{n,r}(u)-u
\right\}^{2}du
=
\frac{n}{3}
+\sum_{i=1}^{n}U_{i,r}^{2}
-\frac{1}{n}\sum_{i=1}^{n}\sum_{j=1}^{n}\max(U_{i,r},U_{j,r})
\end{equation}
for the CCDF test, and $U_{i,r}$ is replaced by $V_{i,r}$ for the
spacings test. The last expression is used for computation and does
not require ordering the transformed values.

If the model parameters were known, the usual CvM reference distribution
for uniformity could be used directly; for example, the $p$-value can be
computed with the function \texttt{cvm.test} in the R package
\texttt{goftest}, which implements the algorithm of
\citet{csorgHo1996exact}. In the present setting, however, the
parameters are estimated from the same data used to construct the
transforms, so the finite-sample null distribution of $W_{n,r}^{2}$ is
affected by parameter estimation. We therefore do not use these
$p$-values directly, and instead calibrate both the CCDF and spacings
tests using a parametric bootstrap.

A fully re-estimated parametric bootstrap requires repeated likelihood
optimization and can be computationally expensive, particularly when the
test is applied sequentially over several values of $r$.
Section~\ref{sec:onestep} develops a one-step parametric-bootstrap
approximation, which closely reproduces the fully re-estimated bootstrap
procedure in our experiments (Section~\ref{sec:mc}).

For each method, the resulting $p$-values are examined sequentially as
$r$ increases. At significance level $\alpha$, the first rejection at
order $r\ge2$ leads to
\[
\widehat r=r-1.
\]
If no rejection occurs up to the prespecified maximum order $R$, we set
$\widehat r=R$.

\section{Theoretical and computational developments}
\label{sec:theory}

This section develops two results underlying the proposed selection
procedures. First, we establish an explicit theoretical relationship
between the CCDF and exponential-spacings statistics. Second, we develop
a one-step parametric-bootstrap approximation that avoids repeated
likelihood optimization.

\subsection{Theoretical relationship between the spacings and CCDF statistics}
\label{sec:comparison}

For $j=1,\ldots,r$, let
\[
Z_j=-\log F_{\theta}(X^{(j)})
=\Bigl\{1-k\frac{X^{(j)}-\mu}{\sigma}\Bigr\}^{1/k},
\qquad
C_j=Z_j-Z_{j-1}\quad (j\ge2).
\]
Under the $r$GEV model, $Z_j\sim\mathrm{Gamma}(j,1)$, and
$C_2,\ldots,C_r$ are independent standard exponential variables
\citep{Weissman1978,tawn1988extreme}. The CCDF value in
Eq.~(\ref{Ur}) and the spacing $S_r$ in Eq.~(\ref{spacing_exp}) can be
written in terms of $Z_r$ and $C_r$ as
\[
U_r=\exp(-C_r),
\qquad
S_r=(r-1)\log\frac{Z_r}{Z_{r-1}}
=(r-1)\log\Bigl(1+\frac{C_r}{Z_{r-1}}\Bigr).
\]
Thus the two tests use the same exponential increment $C_r$, but
through different functions.

\begin{proposition}
	\label{prop:spacing_ccdf}
	Let $r\ge2$, and let $C_r$ and $S_r$ be computed at the true parameter
	$\theta_0$. Then $C_r$ and $S_r$ are standard exponential,
	\begin{equation}
	\label{eq:corr_sc}
	\mathrm{Corr}(S_r,C_r)=1-\frac{1}{r},
	\end{equation}
	\begin{equation}
	\label{eq:mse_sc}
	E\{(S_r-C_r)^2\}=\frac{2}{r},
	\end{equation}
	and, for every $\epsilon>0$,
	\begin{equation}
	\label{eq:prob_sc}
	P(|S_r-C_r|>\epsilon)\le\min\{1,2/(r\epsilon^2)\}.
	\end{equation}
\end{proposition}

The proof, based on the beta--gamma relationship $Z_{r-1}/Z_r\sim\mathrm{Beta}(r-1,1)$,
is given in the Appendix. As $r$ increases, the correlation of the two
transformed quantities approaches one and their mean-square discrepancy
converges to zero. This helps explain the similar $r$-selection behavior of
the two procedures observed in the Monte Carlo experiments.

\subsection{One-step parametric-bootstrap approximation}
\label{sec:onestep}

Because the null distribution of a goodness-of-fit statistic is affected
when model parameters are estimated from the same data, we calibrate the
Cram\'er--von Mises statistic using a parametric bootstrap. The need to
account for parameter estimation in probability-transform-based
goodness-of-fit procedures is well recognized; see, for example,
\citet{GenestQuessyRemillard2006}, \citet{DobricSchmid2007}, and
\citet{BergBakken2007}. Parametric-bootstrap calibration is also standard
more generally for composite goodness-of-fit testing with estimated
parameters
\citep{BurkeGombay1988,Stutetal1993, GenestRemillard2008}.
A closely related parametric-bootstrap construction was used by
\citet{bader2017automated} for the $r$GEV score test. But this procedure requires numerical
optimization for every bootstrap replicate.
Our approach avoids this optimization by replacing each bootstrap MLE
with a one-step Newton update from the original estimate. For regular
models, one-step and $k$-step bootstrap estimators are known to be
first-order asymptotically equivalent to the fully re-estimated
bootstrap estimators \citep{DavidsonMacKinnon1999,Andrews2002}, and
related linearization-based fast bootstraps have been shown to be
consistent \citep{SalibianBarreraZamar2002}. Although these results
were not derived for the $r$GEV model, its MLE is asymptotically normal
when $k<1/2$ \citep{Smith1985}, so similar theoretical support is
expected for the one-step bootstrap considered here.

Let $\widehat{\theta}_n$ denote the MLE under the fitted
$r$GEV model and let $\ell_n(\theta)$ denote the corresponding
log-likelihood in Eq.~(\ref{gevnllh}). A bootstrap sample consists of $n$ blocks
generated independently from $f_r(\cdot;\widehat{\theta}_n)$.
Write $\ell_n^*(\theta)$ for the log-likelihood of a
bootstrap sample and
\[
\boldsymbol g_n^*(\theta)
=
\nabla \ell_n^*(\theta)
\]
for its gradient. Throughout, an asterisk denotes a quantity computed
from a bootstrap sample. The fully re-estimated bootstrap MLE is
\[
\widehat{\theta}_n^*
=
\arg\max_{\theta}
\ell_n^*(\theta).
\]

To avoid repeated optimization, we replace
$\widehat{\theta}_n^*$ by the one-step (OS) approximation
\begin{equation}
\label{eq:onestep}
\widetilde{\theta}_n^*
=
\widehat{\theta}_n
+
\boldsymbol H_n(\widehat{\theta}_n)^{-1}
\boldsymbol g_n^*(\widehat{\theta}_n),
\end{equation}
where
\[
\boldsymbol H_n(\widehat{\theta}_n)
=
-\nabla^2 \ell_n(\widehat{\theta}_n)
\]
is the observed information matrix, that is, the negative Hessian of
the log-likelihood, based on the original sample. The Hessian is
computed only once and reused for all bootstrap replicates. Thus, each
bootstrap parameter estimate requires only evaluation of the bootstrap
gradient and a matrix update rather than a new numerical optimization.
This construction is motivated by classical one-step estimation
\citep{LeCam1956,vanderVaart1998} and is related to the $k$-step
bootstrap of \citet{Andrews2002}.

For numerical implementation, the bootstrap gradient was evaluated
analytically. If the full update in Eq.~(\ref{eq:onestep}) left the
admissible $r$GEV parameter region, we used the step-halved update
\[
\widetilde{\theta}_n^*
=
\widehat{\theta}_n
+
2^{-m}\boldsymbol H_n(\widehat{\theta}_n)^{-1}
\boldsymbol g_n^*(\widehat{\theta}_n)
\]
with the smallest integer $m\ge1$ for which the scale and support
constraints were satisfied. The construction is also
related in spirit to score-based and multiplier-bootstrap methods through
its use of a first-order likelihood-gradient linearization, although here the
linearization is used specifically to approximate the bootstrap MLE
within a parametric-bootstrap scheme.

For a bootstrap sample, let $U_{i,r}^*(\theta)$ and $V_{i,r}^*(\theta)$,
$i=1,\ldots,n$, denote the CCDF and spacings values in Eqs.~(\ref{Ur})
and~(\ref{spacing_unif}), computed from the bootstrap blocks with
parameter $\theta$. Let $F_{n,r}^*(\cdot\,;\theta)$ denote the
empirical distribution function of $U_{1,r}^*(\theta),\ldots,U_{n,r}^*(\theta)$
for the CCDF test, or of $V_{1,r}^*(\theta),\ldots,V_{n,r}^*(\theta)$
for the spacings test, and let
\begin{equation}
\label{eq:cvm_boot}
W_{n,r}^{2*}(\theta)
=n\int_0^1\left\{F_{n,r}^*(u;\theta)-u\right\}^2du,
\end{equation}
which can be computed in the same way as Eq.~(\ref{cvm_stat}).
The one-step and fully re-estimated bootstrap statistics are
$W_{n,r}^{2*}(\widetilde{\theta}_n^*)$ and
$W_{n,r}^{2*}(\widehat{\theta}_n^*)$. Let $W_{n,r}^{2}$ denote the
observed statistic in Eq.~(\ref{cvm_stat}), computed from the original
sample with $\widehat{\theta}_n$. With $B_{\rm boot}$ bootstrap samples,
the OS and full-bootstrap $p$-values are
\[
p_{\rm OS}
=\frac{1}{B_{\rm boot}}\sum_{b=1}^{B_{\rm boot}}
\mathbf{1}\bigl\{W_{n,r,b}^{2*}(\widetilde{\theta}_{n,b}^*)\ge W_{n,r}^{2}\bigr\},
\]
\[
p_{\rm Full}
=\frac{1}{B_{\rm boot}}\sum_{b=1}^{B_{\rm boot}}
\mathbf{1}\bigl\{W_{n,r,b}^{2*}(\widehat{\theta}_{n,b}^*)\ge W_{n,r}^{2}\bigr\},
\]
where the subscript $b$ refers to the $b$th bootstrap sample and
$\mathbf{1}\{\cdot\}$ is the indicator function.

The null distribution of a CvM statistic with estimated parameters
depends on $\theta_0$ \citep{Durbin1973}, which is why bootstrap
calibration is needed. The one-step update replaces the bootstrap MLE by a
single Newton step from $\widehat{\theta}_n$. In finite samples the one-step update may
be less stable when the fitted parameter is close to the boundary of the
admissible support or when a few extreme observations have high leverage,
which the support checks and step-halving described above are intended to
mitigate. Finite-sample accuracy and computational gain are examined
empirically in Section~\ref{sec:mc} by comparison with the fully
re-estimated bootstrap.

\section{Monte Carlo experiments}
\label{sec:mc}

We conducted Monte Carlo experiments to examine finite-sample calibration, power, sequential
selection of $r$, and the consequences of selecting $r$ for extreme
return-level estimation. We also investigated performance under model
misspecification and assessed the accuracy and computational gain of the
one-step parametric-bootstrap approximation developed in
Section~\ref{sec:onestep}.

Unless otherwise stated, the location and scale parameters were fixed at
$\mu=0$ and $\sigma=1$, the nominal significance level was 0.05, and
1,000 Monte Carlo replications were used. For the CCDF and spacings
procedures, parametric-bootstrap calibration was based on the one-step
approximation with $B_{\rm boot}=199$ bootstrap samples.

\subsection{Finite-sample calibration and validation of the one-step bootstrap}
\label{sec:cvm_calibration}

We first examined the finite-sample size of the CCDF and spacings tests
under the null $r$GEV model. This experiment is important because the
uniform transformations used by both procedures depend on parameters
estimated from the same data. Consequently, direct use of the standard
CvM null distribution does not account for the effect of parameter
estimation on the finite-sample null distribution.

Independent samples were generated under
\[
n\in\{30,50,80\},\qquad
k\in\{-0.25,0,0.25\},\qquad
r\in\{2,4,6,8\}.
\]
For each of the 36 settings, 1,000 Monte Carlo samples were generated.
We compared the naive CvM test with the one-step
parametric-bootstrap-calibrated test. Here, the naive CvM test refers
to applying the standard CvM reference distribution directly to the
transformed observations evaluated at the estimated parameters, without
adjusting for the effect of parameter estimation.

Table~\ref{tab:size_summary_os} summarizes the empirical sizes.
The uncalibrated tests were systematically conservative. Averaged over
the 36 settings, the empirical rejection rates were 0.0256 for the CCDF
test and 0.0326 for the spacings test. After one-step bootstrap
calibration, the corresponding rates increased to 0.0477 and 0.0487,
respectively, both close to the nominal level of 0.05.

The mean absolute deviation from the nominal level decreased from
0.0244 to 0.0069 for the CCDF test and from 0.0176 to 0.0047 for the
spacings test, reductions of approximately 72\% and 73\%,
respectively. The improvement was consistent across the sample sizes,
shape parameters, and values of $r$ considered.


\begin{table}[htb]
	\centering
	\caption{Empirical size of the naive and one-step (OS)
		parametric-bootstrap calibrated CvM tests.
		Values are averaged over the 12 combinations of $k$ and $r$
		within each sample size. MAD denotes the mean absolute
		deviation from the nominal level 0.05.}
	\label{tab:size_summary_os}
	\begin{tabular}{c cc cc cc cc}
		\toprule
		& \multicolumn{4}{c}{CCDF}
		& \multicolumn{4}{c}{Spacings}\\
		\cmidrule(lr){2-5}\cmidrule(lr){6-9}
		$n$
		& \multicolumn{2}{c}{Naive}
		& \multicolumn{2}{c}{OS}
		& \multicolumn{2}{c}{Naive}
		& \multicolumn{2}{c}{OS}\\
		& Size & MAD & Size & MAD
		& Size & MAD & Size & MAD\\
		\midrule
		30 & 0.0259 & 0.0241 & 0.0477 & 0.0068
		& 0.0324 & 0.0176 & 0.0484 & 0.0038\\
		50 & 0.0264 & 0.0236 & 0.0484 & 0.0083
		& 0.0327 & 0.0178 & 0.0491 & 0.0066\\
		80 & 0.0244 & 0.0256 & 0.0470 & 0.0057
		& 0.0326 & 0.0174 & 0.0487 & 0.0038\\
		\midrule
		Overall
		& 0.0256 & 0.0244 & 0.0477 & 0.0069
		& 0.0326 & 0.0176 & 0.0487 & 0.0047\\
		\bottomrule
	\end{tabular}
\end{table}

Figure~S3 in the Supplementary Information displays the empirical
rejection rates for all 36 settings. The naive CCDF test is particularly conservative
for smaller values of $r$, whereas the bootstrap-calibrated test is
generally close to the nominal level. A similar, although somewhat less
pronounced, pattern is observed for the spacings test. Occasional mild
over-rejection remains after calibration, but the deviations are small
relative to the conservativeness of the uncalibrated procedures.

To evaluate the one-step approximation itself, we additionally compared
it with a fully re-estimated parametric bootstrap in five representative
settings. The same bootstrap samples were used for the two procedures
within each Monte Carlo replication, so that differences arose only
from the parameter re-estimation step.

The calibrated $p$-values from the two bootstrap procedures were very
close. The mean absolute difference was 0.0099 for the CCDF test and
0.0109 for the spacings test, and their 5\% testing decisions agreed in
98.8\% and 99.2\% of the replications, respectively. Across the ten
test-by-setting comparisons, the correlations between the one-step and
full-bootstrap $p$-values ranged from 0.9973 to 0.9995.


\begin{table}[htb]
	\centering
	\caption{Validation of the one-step parametric bootstrap against the
		fully re-estimated parametric bootstrap in five representative settings.
		For each setting, 100 Monte Carlo samples and 199 bootstrap samples were
		used.  $\Delta_p$ denotes the mean absolute difference
		$|p_{\rm OS}-p_{\rm Full}|$, and Agree denotes the proportion of identical
		5\% testing decisions. OS and Full denote the one-step and fully
		re-estimated parametric-bootstrap calibrations.}
	\label{tab:os_full_validation}
	\setlength{\tabcolsep}{5pt}
	\begin{tabular}{cccccccc}
		\toprule
		$n$ & $k$ & $r$ & Test & Size$_{\rm OS}$ & Size$_{\rm Full}$
		& $\Delta_p$ & Agree \\
		\midrule
		30 & $-0.25$ & 2 & CCDF     & 0.04 & 0.04 & 0.0159 & 0.96 \\
		&         &   & Spacings & 0.05 & 0.04 & 0.0168 & 0.99 \\
		30 & $0.25$ & 4 & CCDF     & 0.04 & 0.04 & 0.0098 & 1.00 \\
		&        &   & Spacings & 0.04 & 0.04 & 0.0098 & 1.00 \\
		30 & $0$    & 8 & CCDF     & 0.05 & 0.05 & 0.0099 & 1.00 \\
		&        &   & Spacings & 0.04 & 0.04 & 0.0106 & 1.00 \\
		50 & $0$    & 4 & CCDF     & 0.04 & 0.04 & 0.0075 & 0.98 \\
		&        &   & Spacings & 0.05 & 0.07 & 0.0098 & 0.98 \\
		80 & $0.25$ & 8 & CCDF     & 0.04 & 0.04 & 0.0067 & 1.00 \\
		&        &   & Spacings & 0.03 & 0.04 & 0.0076 & 0.99 \\
		\midrule
		\multicolumn{4}{r}{Average $\Delta_p$ / agreement: CCDF}
		& \multicolumn{2}{c}{}
		& 0.0099 & 0.988 \\
		\multicolumn{4}{r}{Average $\Delta_p$ / agreement: Spacings}
		& \multicolumn{2}{c}{}
		& 0.0109 & 0.992 \\
		\bottomrule
	\end{tabular}
\end{table}

The computational saving was substantial. The bootstrap re-estimation
step was approximately 65--146 times faster with the one-step update
than with full likelihood maximization. Including the common costs of
the original fit, Hessian evaluation, transformation, and bootstrap
sample generation, target-$r$ calibration was approximately 25--34
times faster. Moreover, the one-step procedure was numerically stable. These results support the one-step procedure as an
accurate and computationally efficient approximation to the full
parametric bootstrap in the settings considered here.

\subsection{Power}
\label{sec:mc_power}

We next assessed the power of the individual goodness-of-fit tests
separately from the sequential selection procedure. Following the
simulation design of \cite{bader2017automated}, we generated the top
seven order statistics from the $7$GEV distribution and replaced the
sixth order statistic by
\begin{equation}
\label{eq:power_mix}
X^{(6),c}
=
pX^{(6)}+(1-p)X^{(7)}.
\end{equation}
The tests were then applied to the top six order statistics, so that
$H_0^{(6)}$ was the target hypothesis.

We considered
\[
n=30,50,80,\qquad
k=-0.2,0,0.2,
\]
and $p=0.8$ and $0.5$. Because the contaminated observation moves
closer to $X^{(7)}$ as $p$ decreases, $p=0.5$ represents the stronger
departure from the null model. The CCDF and spacings tests used the
one-step bootstrap calibration with $B_{\rm boot}=199$.

Table~\ref{tab:power_r6} shows the empirical power pooled over the three
shape parameters. The CCDF and spacings tests exhibit very similar
power. Under the milder departure $p=0.8$, their rejection
probabilities increase from 0.235 and 0.218 at $n=30$ to 0.597 and
0.563 at $n=80$, respectively. Under the stronger departure $p=0.5$,
both procedures have power exceeding 0.80 at $n=30$ and approach one
for $n\ge50$.

The ED test also gains power as the departure becomes stronger and the
sample size increases, but is less sensitive to this particular
alternative. Results for individual values of $k$ are provided in the
Supplementary Information.

\begin{table}[htb]
	\centering
	\caption{Empirical power for testing $H_0^{(6)}$ under the
		Bader-type mixture alternative. The reported values are pooled over
		$k=-0.2,0,$ and $0.2$ by dividing the total number of rejections by
		the total number of valid replications. Smaller $p$ corresponds to a
		stronger departure from the null.}
	\label{tab:power_r6}
	\setlength{\tabcolsep}{7pt}
	\begin{tabular}{ccccc}
		\toprule
		$n$ & $p$ & CCDF & Spacings & ED \\
		\midrule
		30 & 0.8 & 0.235 & 0.218 & - \\
		30 & 0.5 & 0.809 & 0.809 & - \\
		\midrule
		50 & 0.8 & 0.478 & 0.495 & 0.107 \\
		50 & 0.5 & 0.980 & 0.967 & 0.653 \\
		\midrule
		80 & 0.8 & 0.597 & 0.563 & 0.130 \\
		80 & 0.5 & 0.997 & 0.997 & 0.887 \\
		\bottomrule
	\end{tabular}
\end{table}

\subsection{Sequential selection under the $r$GEV model}
\label{sec:mc_rgev}

We next examined sequential selection of $r$ under the $r$GEV
framework. Five shape parameters were considered:
\[
k=-0.35,-0.2,0,0.2,0.35,
\]
with sample sizes $n=30,50,$ and $80$. The top ten order statistics
were initially generated from the $10$GEV model using the algorithm of
\cite{bader2017automated}.

To create a known departure point, the sixth through ninth order
statistics were modified according to
\begin{equation}
\label{eq:sequential_mix}
X^{(j),c}
=
pX^{(j)}+(1-p)X^{(j+1)},
\qquad j=6,\ldots,9.
\end{equation}
Thus, the largest correctly specified order, denoted by $r_0$, is $r_0=5$.
We considered
$p=0.5$ and $0.8$, with $p=0.5$ representing the stronger departure,
and set the maximum candidate order to $R=10$.

\begin{figure}[h!]
	\vspace{0.3cm}
	\centering
	\includegraphics[width=13cm,height=9cm]{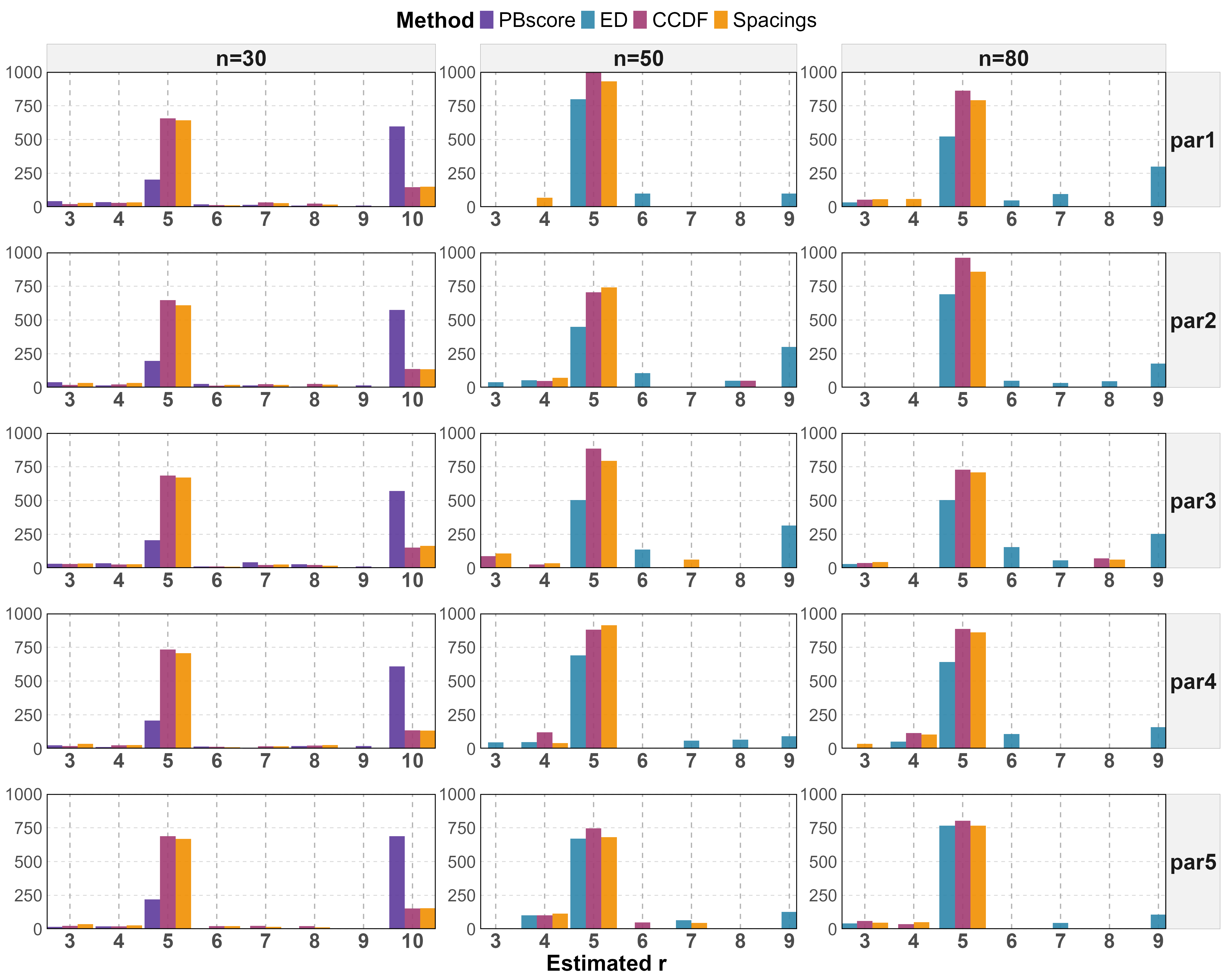}
	\caption{Histograms of the selected $r$ for five $r$GEV parameter settings
		with largest correctly specified order $r_0=5$, based on 1,000
		Monte Carlo samples with $n=30,\,50$, and $80$ and mixing probability
		$p=0.5$. The shape parameters from top to bottom are
		par1: $k=-0.35$, par2: $k=-0.2$, par3: $k=0$,
		par4: $k=0.2$, and par5: $k=0.35$.}
	\label{fig:rsel_sim_gev}
\end{figure}

Figure~\ref{fig:rsel_sim_gev} presents the distributions of the selected
$r$ for $p=0.5$. The CCDF and spacings procedures show very similar
selection behavior, consistent with the theoretical relationship
established in Proposition~\ref{prop:spacing_ccdf}. Both tend to select
orders near $r_0=5$, with the selection distributions becoming more
concentrated around $r_0$ as the sample size increases. The overall
pattern is broadly similar across the five shape parameters.

For comparison with \cite{bader2017automated}, the score test was
included for $n=30$, whereas the ED test was used for $n\ge50$.
The score test performs relatively poorly in the small-sample setting
in these experiments. For $n\ge50$, the ED procedure is competitive
but tends to show somewhat less concentrated selection around $r_0=5$
than the CCDF and spacings procedures in these experiments.

Results for the milder departure $p=0.8$ are given in the
Supplementary Information. As expected, distinguishing the departure
from the $r$GEV model becomes more difficult when the contamination is
weaker.

\subsection{Performance under model misspecification}
\label{sec:mc_wakeby}

To examine sensitivity to distributional misspecification, we
generated samples from the Wakeby distribution rather than from an
$r$GEV model. The five-parameter Wakeby family can represent a broad
range of distributional shapes and has been described as a
``mimic-everything'' distribution
\citep{houghton1978birth,LandwehrMatalasWallis1979Wakeby,
	hosking1997regional}.

We considered the six Wakeby distributions WA-1 through WA-6 of
\cite{LandwehrMatalasWallis1979Wakeby}. Their parameter settings,
expressed using the reparametrization of \cite{hosking1997regional},
are reported in the Supplementary Information, together with the
algorithm used to generate the rLOS samples.

For this experiment, each block consisted of
$B_{\rm block}=100$ observations generated from the specified Wakeby
distribution, from which the largest order statistics were extracted.
For each distribution, the top ten order statistics were generated and
the sixth through ninth were modified according to
Eq.~(\ref{eq:sequential_mix}). Since the underlying distribution is not
$r$GEV, $r_0=5$ should here be interpreted as the designated onset of
the artificial departure rather than as a true $r$GEV order.

\begin{figure}[h!]
	\vspace{0.3cm}
	\centering
	\includegraphics[width=13cm,height=9cm]{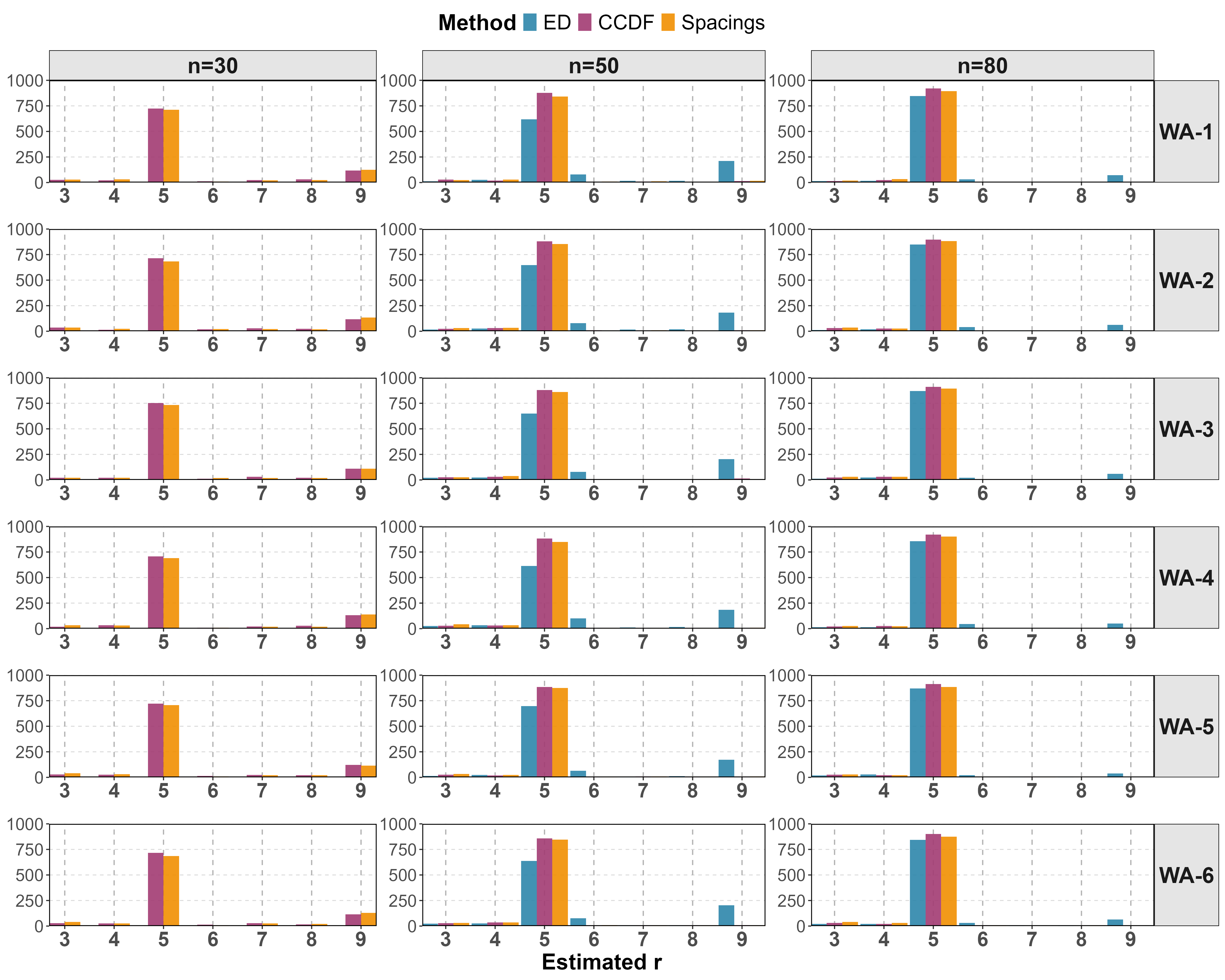}
	\caption{Histograms of the selected $r$ for the six Wakeby
			distributions with designated change point $r_0=5$ and mixing
			probability $p=0.5$. 
		The parameter settings from top to bottom correspond to WA-1 through WA-6;
		details are provided in the Supplementary Information. Other simulation
		settings are the same as those in Figure~\ref{fig:rsel_sim_gev}.}
	\label{fig:rsel_sim_wakeby}
\end{figure}

Figure~\ref{fig:rsel_sim_wakeby} presents the selected-$r$
distributions for $p=0.5$. The CCDF and spacings procedures again show
similar behavior and generally identify orders near the designated
change point. For $n\ge50$, the ED test tends to be somewhat less
concentrated around $r_0=5$. The score test was omitted because of its
substantially greater computational cost and its comparatively weak
performance in the preceding experiments.


These results indicate that the CCDF and spacings procedures retain
useful selection behavior under the forms of marginal misspecification
considered here. Results for the milder departure $p=0.8$ are provided
in the Supplementary Information.

Additional sensitivity experiments, reported in the Supplementary
Information, examined a Wakeby baseline without the artificial
contamination, block sizes $B_{\rm block}=50,100,$ and $365$, and
within-block dependence generated by a Gaussian AR(1) latent process
with Wakeby margins. The CCDF and spacings procedures again behaved
similarly, block size had no systematic monotone effect, and strong
within-block dependence (lag-one correlation $0.6$ of the latent
Gaussian process) led to earlier rejection.

\subsection{Consequences for return-level estimation}
\label{sec:rl_consequence}

Goodness-of-fit-based selection of $r$ is ultimately useful only insofar
as it supports the inferential objective of the analysis. We therefore
examined the effect of order selection on estimation of the 100-year
return level.

The data-generating mechanism was the same as in
Section~\ref{sec:mc_rgev} with $p=0.5$ and largest correctly specified
order $r_0=5$. We considered
\[
n=30,80,\qquad k=-0.2,0,0.2,
\]
with 500 Monte Carlo replications for each setting.

We compared the block-maxima estimator ($r=1$), an oracle-$r_0$
estimator using the known largest correctly specified order $r_0=5$,
and estimators based on the orders selected by the CCDF and spacings
procedures. For $n=80$, we additionally included the estimator based on
the order selected by the ED procedure. Unlike the usual oracle
estimator, the oracle-$r_0$ estimator uses knowledge of the
model-validity boundary $r_0$, not of the RMSE-minimizing order.
The CCDF and spacings procedures used $B_{\rm boot}=199$, with selection
based on the first rejection of the sequential unadjusted $p$-value path
at the 5\% level.

Table~\ref{tab:rl100_main} reports the bias, standard deviation, and RMSE of the
estimated 100-year return level. In all six settings, the oracle-$r_0$,
CCDF-selected, and spacings-selected estimators have smaller RMSE than
the block-maxima estimator. For the oracle $r_0=5$ fit, the RMSE
reduction relative to block maxima ranges from approximately 12\% to
29\%. The improvement is mainly attributable to reduced sampling
variability rather than to a systematic reduction in bias.

\begin{table}[!htb]
	\centering
	\small
	\caption{Bias, standard deviation (SD), and RMSE of the estimated
		100-year return level. The oracle uses the known largest correctly
		specified order $r_0=5$. BM denotes the block-maxima ($r=1$)
		estimator. CCDF-OS and Spacings-OS denote the CCDF and
		spacings procedures calibrated by the one-step (OS) parametric
		bootstrap. ED is reported only for $n=80$.
		Each setting is based on 500 Monte Carlo replications.}
	\label{tab:rl100_main}
	\setlength{\tabcolsep}{5pt}
	\begin{tabular}{ccrrrrr}
		\toprule
		$n$ & $k$ & BM & Oracle & CCDF-OS & Spacings-OS & ED \\
		\midrule
		
		\multicolumn{7}{l}{\textit{Panel A: Bias}}\\
		30 & $-0.2$ &  0.272 &  0.231 &  0.273 &  0.313 & -- \\
		30 & $ 0$   & -0.011 & -0.063 &  0.023 &  0.005 & -- \\
		30 & $ 0.2$ & -0.111 & -0.131 & -0.107 & -0.107 & -- \\
		80 & $-0.2$ &  0.120 &  0.040 &  0.063 &  0.055 &  0.079 \\
		80 & $ 0$   & -0.016 & -0.001 & -0.004 & -0.001 &  0.004 \\
		80 & $ 0.2$ & -0.058 & -0.037 & -0.039 & -0.040 & -0.036 \\
		
		\midrule
		\multicolumn{7}{l}{\textit{Panel B: SD}}\\
		30 & $-0.2$ & 3.464 & 2.745 & 2.712 & 2.736 & -- \\
		30 & $ 0$   & 1.605 & 1.221 & 1.309 & 1.268 & -- \\
		30 & $ 0.2$ & 0.785 & 0.552 & 0.655 & 0.564 & -- \\
		80 & $-0.2$ & 2.136 & 1.578 & 1.650 & 1.634 & 1.560 \\
		80 & $ 0$   & 0.900 & 0.733 & 0.752 & 0.763 & 0.725 \\
		80 & $ 0.2$ & 0.359 & 0.318 & 0.326 & 0.326 & 0.328 \\
		
		\midrule
		\multicolumn{7}{l}{\textit{Panel C: RMSE}}\\
		30 & $-0.2$ & 3.475 & 2.755 & 2.726 & 2.754 & -- \\
		30 & $ 0$   & 1.605 & 1.223 & 1.309 & 1.268 & -- \\
		30 & $ 0.2$ & 0.793 & 0.567 & 0.664 & 0.574 & -- \\
		80 & $-0.2$ & 2.139 & 1.579 & 1.651 & 1.635 & 1.562 \\
		80 & $ 0$   & 0.900 & 0.733 & 0.752 & 0.763 & 0.725 \\
		80 & $ 0.2$ & 0.364 & 0.320 & 0.328 & 0.328 & 0.330 \\
		
		\bottomrule
	\end{tabular}
\end{table}

The reduction in RMSE is driven primarily by lower sampling variability.
For example, at $n=30$ and $k=-0.2$, the SD decreases from 3.464 for
the block-maxima estimator to 2.745 for the oracle and 2.712--2.736
for the two data-driven procedures, while the corresponding biases
remain relatively small compared with the sampling variability.
Similar variance reductions are evident across the other settings.

\begin{figure}[htb]
	\centering
	\includegraphics[width=14cm]{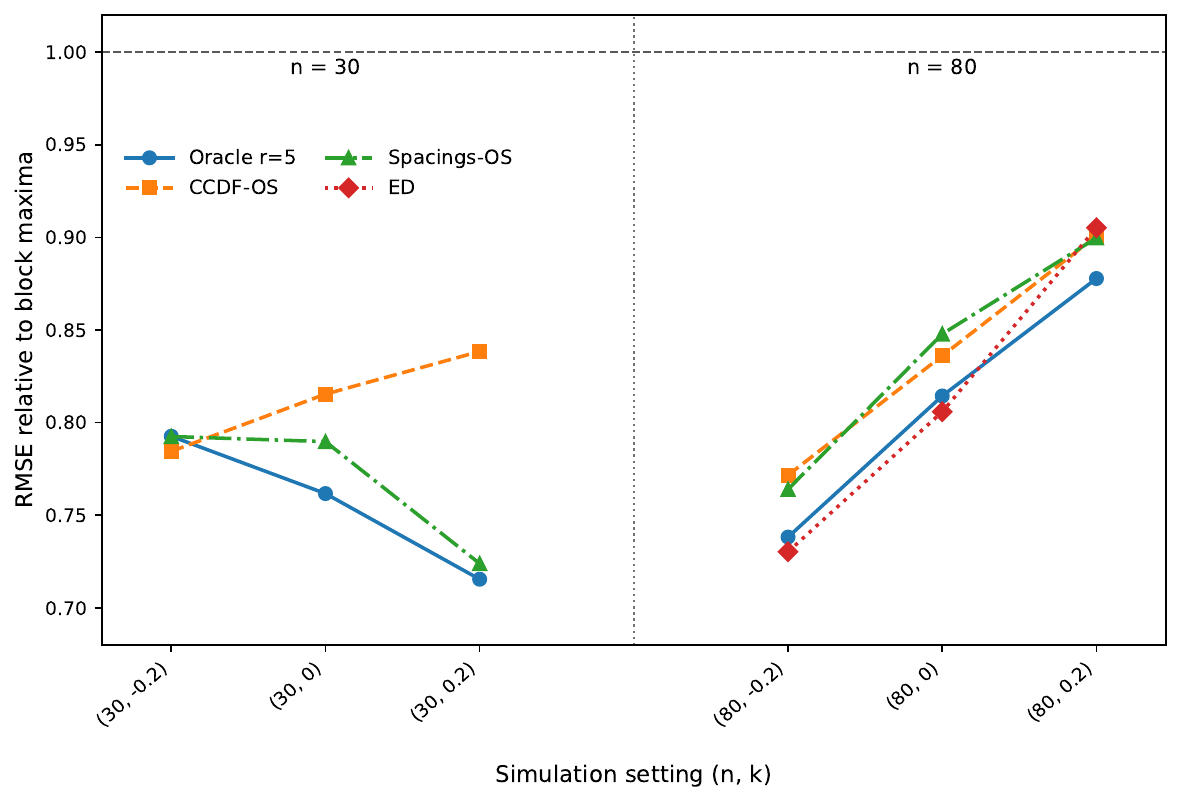}
	\caption{RMSE of the estimated 100-year return level relative to the
		block-maxima ($r=1$) estimator. Values below one indicate lower RMSE than
		the block-maxima fit. The oracle fit uses the known largest correctly specified order
		$r_0=5$.}
	\label{fig:rl100_relative_rmse}
\end{figure}

Figure~\ref{fig:rl100_relative_rmse} shows RMSE relative to the
block-maxima estimator. The CCDF and spacings procedures generally
retain a substantial portion of the efficiency gain obtained by using
the additional upper order statistics. For $n=80$, the ED procedure
shows similarly low RMSE.


To investigate this point directly, we conducted an additional
fixed-$r$ experiment. Under the same six settings, the $r$GEV model
was fitted separately at each fixed value $r=1,\ldots,10$, and the bias, standard deviation, and RMSE of the estimated 100-year
return level were calculated.

The RMSE-minimizing orders were $r^\star=(6,8,9)$ for $n=30$ and
$k=(-0.2,0,0.2)$, respectively, and $r^\star=(6,7,9)$ for $n=80$. Thus, in all six settings the finite-sample
RMSE-minimizing order exceeded the largest correctly specified order
$r_0=5$. Relative to the $r=5$ estimator, the reductions in RMSE at
$r^\star$ ranged from approximately 2.9\% to 13.7\%.

This behavior illustrates the bias--variance trade-off inherent in the
choice of $r$. Moderate inclusion of lower order statistics beyond
$r_0$ introduces some model misspecification, but the associated
reduction in sampling variance can more than compensate for the
resulting bias. The benefit does not continue indefinitely: by $r=10$,
the RMSE exceeded that at $r=5$ in all six settings.

For reference, Tables~S6 and~S7 in the Supplementary Information show that a fixed order moderately above
$r_0$ (e.g., $r=8$) has lower RMSE than the CCDF- and spacings-selected
estimators in all six settings. Such an order, however, cannot be
identified in advance, since the RMSE-minimizing $r$ depends on the
unknown data-generating mechanism (Table~S8). These results indicate room for
improvement of inference-oriented selection rather than a competing rule.
Figure~S4 in the Supplementary Information illustrates the RMSE
trajectories relative to the $r=5$ fit. Detailed fixed-$r$ results for bias,
standard deviation, and RMSE are reported in the Supplementary
Information.

\section{Real-data application}
\label{sec:application}

We illustrate the proposed selection procedures using daily extreme
rainfall observations from Sancheong, Korea (station ID 289). For each
year from 1973 to 2022, the 20 largest daily precipitation amounts were
extracted, yielding $n=50$ annual blocks with a maximum candidate order
of $R=20$. The precipitation amounts are measured in millimeters and
were obtained from the Korea Meteorological Administration.

Figure~S5 in the Supplementary Information presents histograms,
time-series plots, and pairwise scatterplots of the first five annual
order statistics. Mann--Kendall tests provided no significant evidence
of monotonic temporal trends in these series. We therefore use the
stationary $r$GEV model as a working specification for this application.

\subsection{Selection of $r$ for the $r$GEV model}

Table~\ref{tab:sancheong_results} reports the parameter estimates,
estimated 100-year return levels (RLs), their standard errors (SEs), and
the $p$-values of the ED, CCDF, and spacings tests as $r$ increases from
1 to 12. The return-level SEs were obtained by the delta method
\citep{coles2001introduction}.

The CCDF and spacings tests were calibrated using the one-step
parametric bootstrap with $B_{\rm boot}=1999$. Their $p$-values remain
above 0.05 through $r=7$ and first fall below 0.05 at $r=8$, with
$p=0.0255$ and $0.0180$, respectively. Thus, both procedures select
\[
\widehat r=7.
\]
The fully re-estimated parametric bootstrap gives the same first
rejection at $r=8$, with corresponding $p$-values 0.0175 and 0.0155.
The ED test first rejects at $r=7$ ($p=0.0497$) and therefore selects
$\widehat r=6$. Since selection is based on the first rejection,
$p$-values at subsequent orders do not affect the selected value of
$r$.

\begin{table}[!htb]
	\centering
	\small
	\caption{Parameter estimates, estimated 100-year return levels
		(RLs), their standard errors (SEs), and sequential-test
		$p$-values for the Sancheong rainfall data. The CCDF and
		spacings $p$-values are calibrated using the one-step
		parametric bootstrap with $B_{\rm boot}=1999$. Sequential
		testing was carried out up to $R=20$; results through $r=12$
		are displayed. The ED test is defined for $r\ge2$; the $r=1$ entries for the CCDF
		and spacings columns correspond to the common ordinary GEV
		probability-transform diagnostic.}
	\label{tab:sancheong_results}
	\setlength{\tabcolsep}{4.5pt}
	\begin{tabular}{rrrrrrrrr}
		\toprule
		$r$ & Loc. & Scale & Shape & RL & SE
		& $p_{\rm ED}$ & $p_{\rm CCDF}$ & $p_{\rm Spac}$ \\
		\midrule
		1  & 134.9 & 56.83 &  0.041 & 372.8 & 66.4
		& -- & 0.5620 & 0.5620 \\
		2  & 137.1 & 55.63 & -0.023 & 407.2 & 54.6
		& 0.8336 & 0.2275 & 0.0505 \\
		3  & 133.1 & 52.55 & -0.147 & 478.3 & 59.8
		& 0.3479 & 0.3090 & 0.2040 \\
		4  & 131.8 & 52.67 & -0.220 & 551.1 & 63.5
		& 0.4035 & 0.0700 & 0.0650 \\
		5  & 133.9 & 52.38 & -0.179 & 508.2 & 47.0
		& 0.8672 & 0.8515 & 0.5105 \\
		6  & 133.8 & 52.43 & -0.192
		& \underline{521.6} & \underline{43.0}
		& 0.6281 & 0.3170 & 0.8315 \\
		7  & 132.7 & 52.76 & -0.231
		& \underline{565.3} & \underline{43.9}
		& \textbf{0.0497} & 0.1905 & 0.1820 \\
		8  & 132.2 & 53.61 & -0.267 & 617.4 & 45.8
		& 0.0494 & \textbf{0.0255} & \textbf{0.0180} \\
		9  & 132.2 & 52.80 & -0.257 & 596.6 & 39.6
		& 0.6455 & 0.5580 & 0.8715 \\
		10 & 131.9 & 53.70 & -0.279 & 633.7 & 40.0
		& 0.6749 & 0.4845 & 0.3265 \\
		11 & 131.8 & 51.82 & -0.249 & 578.5 & 32.4
		& 0.5752 & 0.8405 & 0.4750 \\
		12 & 131.9 & 51.47 & -0.240 & 564.1 & 29.1
		& 0.4009 & 0.8260 & 0.4665 \\
		\bottomrule
	\end{tabular}
\end{table}

\begin{figure}[h!tb]
	\vspace{0.5cm}
	\centering
	\includegraphics[width=12cm,height=7cm]
	{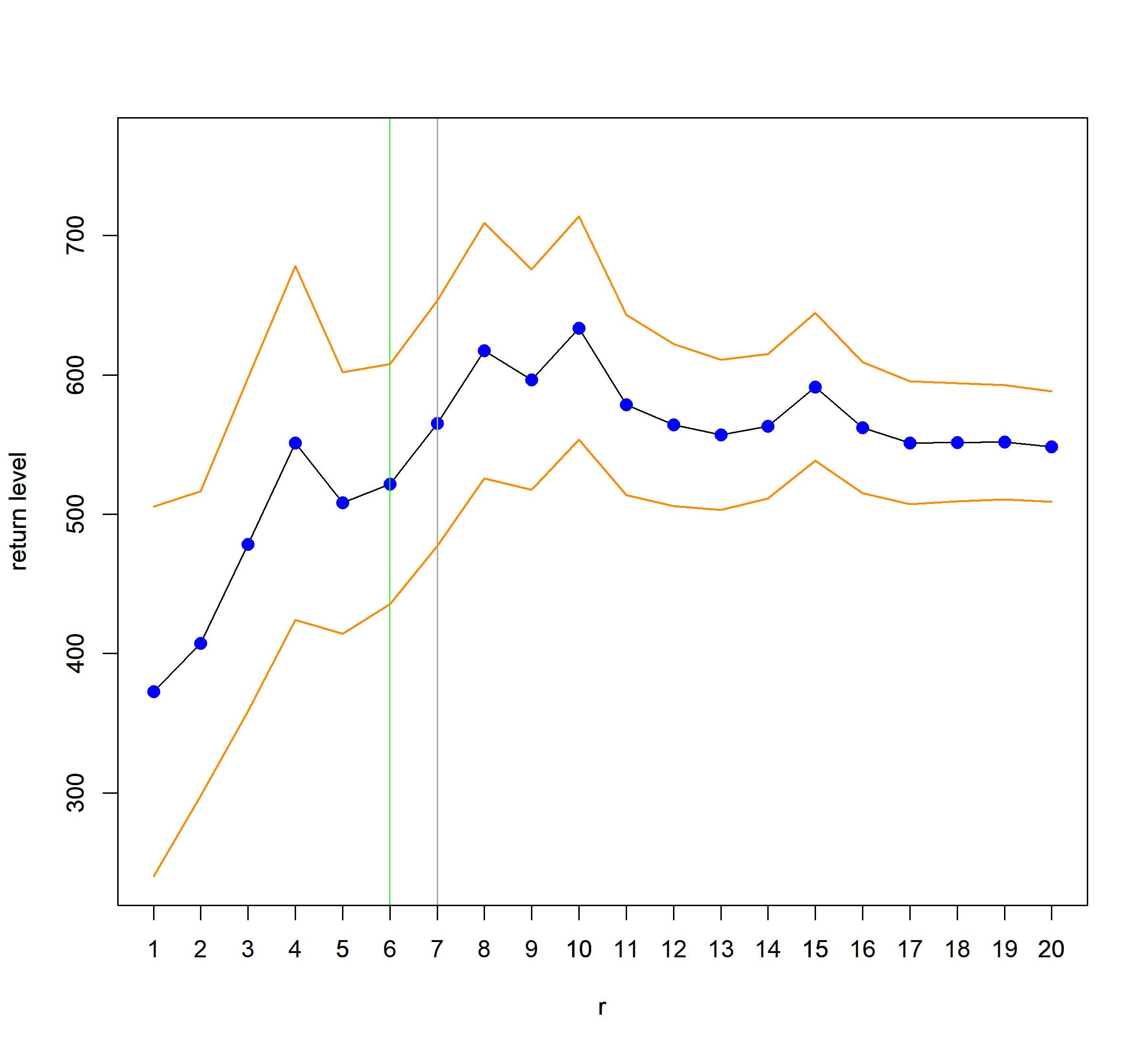}
	\caption{Estimated 100-year return levels and 95\% confidence
		intervals as $r$ increases from 1 to 20 for the Sancheong
		extreme rainfall data.}
	\label{fig:seq_rt_plot}
\end{figure}

Figure~\ref{fig:seq_rt_plot} shows the estimated 100-year return levels
and their 95\% confidence intervals as $r$ increases from 1 to 20. The
confidence intervals were obtained using the delta method, as implemented
in the \texttt{gevrRl} function of the R package \texttt{eva}. The
estimated return level changes noticeably as additional order statistics
are included, while its standard error generally decreases.

At the orders selected by the sequential procedures, the estimated
100-year return levels are 521.6 mm for $r=6$ and 565.3 mm for $r=7$,
with SEs of 43.0 and 43.9 mm, respectively. By comparison, the
block-maxima fit ($r=1$) gives an estimated return level of 372.8 mm
with an SE of 66.4 mm. Thus, inclusion of additional upper order
statistics substantially reduces estimation uncertainty, while the
change in the point estimate illustrates why the choice of $r$ can also
have an important effect on extreme-quantile inference.

Figure~\ref{fig:sancheong_ccdf_pp} and Figure~S6 in the
Supplementary Information provide the corresponding conditional
probability plots through the first rejected order. For both procedures,
the one-step bootstrap $p$-values remain above 0.05 through $r=7$,
whereas the tests reject at $r=8$, with $p=0.0255$ for the CCDF
procedure and $p=0.0180$ for the spacings procedure. Thus, the conditional probability plots complement the sequential
testing results, under which both procedures select $\widehat r=7$.

\begin{figure}[h!tb]
	\vspace{0.5cm}
	\centering
	\includegraphics[width=15cm,height=8.5cm]{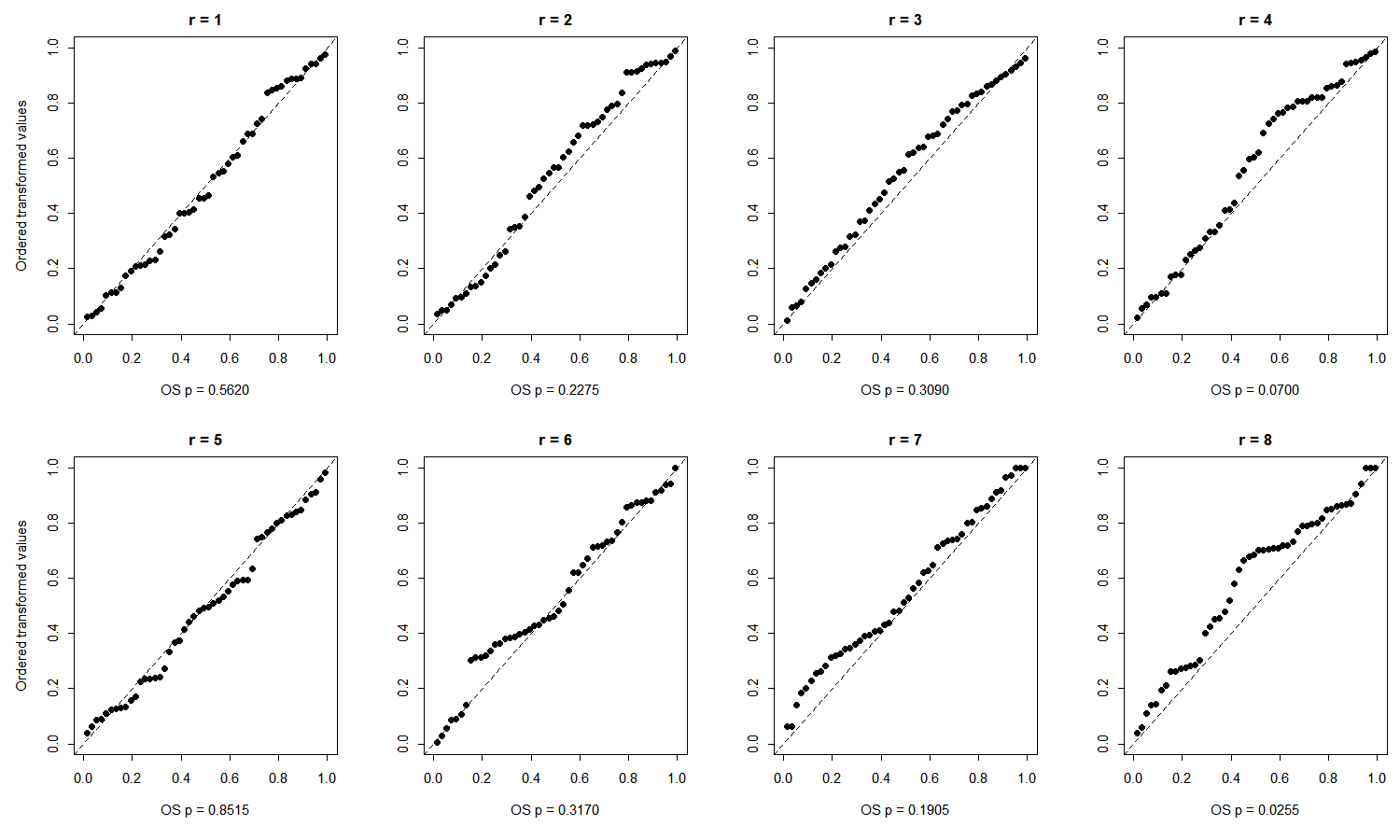}
\caption{Conditional probability plots for $r=1,\ldots,8$ for the
	Sancheong extreme rainfall data based on the CCDF transformation.
	Each panel plots the sorted transformed values against the
	corresponding quantiles of the standard uniform distribution.
	The dashed line is the 45$^\circ$ reference line for the standard
	uniform distribution. The one-step parametric-bootstrap (OS)
	$p$-value is shown below each panel.}
\label{fig:sancheong_ccdf_pp}
\end{figure}

As an additional illustration of the applicability of the conditional
approach beyond the $r$GEV model, we also applied the CCDF procedure
to the four-parameter kappa $r$-largest model of \citet{shin2023modeling}.
The model formulation,
selection results, and diagnostic plots are provided in the
Supplementary Information.

\section{Discussion}
\label{sec:discussion}

The simulation results show that the CCDF and exponential-spacings
procedures have very similar finite-sample behavior. This agreement is
not accidental: Proposition~\ref{prop:spacing_ccdf} establishes an
explicit relationship between their underlying transformed quantities,
with their correlation approaching one and their mean-square discrepancy
decreasing as $r$ increases. The two procedures nevertheless arise from different
constructions. The spacings test exploits a particular exponential-spacing
property of the $r$-largest Gumbel representation, whereas the CCDF
procedure is based directly on the conditional distribution of each newly
included order statistic. The latter formulation provides a natural route
to other rLOS models whenever the corresponding conditional distribution
is available, as illustrated by the four-parameter kappa model in the
Supplementary Information.

The conditional construction also provides a useful graphical diagnostic.
Earlier work \citep{smith1986extreme,tawn1988extreme} considered
probability plots based on the marginal distribution of the $r$th order
statistic. We examined a sequential CvM version of this marginal
diagnostic, but found relatively low power in our simulation experiments.
In contrast, the sequential plots used here focus on the newly added
order statistic: the CCDF plot examines it through its conditional
distribution, and the spacings plot examines it through the corresponding
exponential-spacing property. They therefore complement, rather than replace, conventional
marginal probability plots.

An important finding of the present study is that goodness-of-fit-based
selection of $r$ and inference-oriented selection of $r$ need not have
the same objective. Sequential testing seeks to identify the largest
order that remains compatible with the assumed rLOS model. For estimation
of a particular extreme quantile or return level, however, finite-sample
accuracy is governed by a bias--variance trade-off
(Section~\ref{sec:rl_consequence}). Thus, the order selected by a
goodness-of-fit procedure should not be interpreted as necessarily
optimal for every inferential target.

Two limitations of the return-level study should be noted. It is
restricted to point estimation of a single 100-year return level under
one contamination design, without other values of $r_0$ or
peaks-over-threshold competitors. Moreover, coverage after selection was
not evaluated: the delta-method intervals in Section~\ref{sec:application}
treat the selected $r$ as fixed and ignore the variability of the
selection step, which is left for future work.

Another practical issue is dependence among observations within a block.
Environmental extremes can exhibit temporal clustering, which may affect
the adequacy of the $r$GEV approximation when several upper order
statistics are retained. The sensitivity experiment reported in the
Supplementary Information showed relatively little effect of moderate
dependence in the WA-3 setting, whereas stronger dependence led to
appreciably smaller selected orders. Although this limited experiment
is not intended as a general robustness assessment, it indicates that
substantial within-block clustering can affect automated $r$ selection.
Event-based declustering may therefore be appropriate in applications
where several upper order statistics can originate from the same event
or cluster \citep{tawn1988extreme,FerroSegers2003}. For the Sancheong data,
however, the available data consist of annual ordered values rather than
the complete dated daily sequence, so a reliable event-based declustering
analysis cannot be performed. The application should therefore be viewed
primarily as an illustration of the proposed selection procedures.
A systematic treatment of dependence and declustering within automated
$r$ selection remains an important topic for future study.
Developing an error-control procedure tailored to the dependence
structure of sequential rLOS tests remains an open problem.

A formal asymptotic justification of the
one-step bootstrap is left for future work. Because the one-step
estimator is an explicit linear function of the bootstrap score, its
first-order validity could be derived from the asymptotic linearity of
the MLE, the consistency of the observed information, and a smoothness
condition on the probability transforms, by the general
parametric-bootstrap theory of \citet{GenestRemillard2008}. 

Finally, the CCDF framework can be extended beyond the stationary
$r$GEV model. In nonstationary settings, where location, scale, or
other parameters depend on time or covariates, the same conditional
probability-transform principle can in principle be used, although
additional work is needed for parameter estimation and bootstrap
calibration. 

\section{Conclusion}
\label{sec:conclusion}

We developed an automated sequential procedure for selecting $r$ in
$r$-largest order statistics models using the conditional cumulative
distribution function (CCDF), and revisited the classical
exponential-spacings diagnostic within the same framework. Both
procedures were formulated as Cram\'er--von Mises tests of uniformity.
We established an explicit theoretical relationship between their
underlying transformed quantities, showing that their correlation
approaches one and their mean-square discrepancy decreases as $r$
increases.

To obtain reliable finite-sample calibration without the computational
cost of repeated likelihood maximization, we developed a one-step
parametric-bootstrap approximation. Its formal asymptotic justification
is left for future work. Monte Carlo
experiments showed that bootstrap calibration substantially improves
finite-sample size accuracy, while the CCDF and spacings procedures
exhibit similar selection behavior and good power against the
alternatives considered. The one-step approximation also closely
reproduced the fully re-estimated bootstrap at a substantially lower
computational cost.

 We emphasize that the primary sequential rule is a screening procedure: it does not control the overall error rate, and its behavior when the $r$GEV model holds at all candidate orders, so that the whole sequential path is under the null, was not examined by simulation; Section~\ref{sec:cvm_calibration} examines only the size of the individual tests.

The methodology was
illustrated using Sancheong rainfall data. A supplementary application
to the four-parameter kappa $r$-largest model \citep{shin2023modeling}
further illustrates that
the conditional-distribution framework can extend beyond the $r$GEV
family. Overall, the proposed CCDF test provides a computationally efficient
approach to the automated selection of $r$.

\begin{small}

\renewcommand{\baselinestretch}{1.0}\selectfont
\section*{Statements and Declarations}

\subsection*{Code and data availability}
The R code and annual $r$-largest daily rainfall data from Sancheong,
Korea, used in this study are available in the GitHub repository at
\url{https://github.com/Pjihong/select-r-ccdf.git}.

\subsection*{ORCID}
Yire Shin: 0000-0003-1297-5430; ~~~
Jihong Park: 0009-0003-3191-9968; \\
Jeong-Soo Park: 0000-0002-8460-4869.

\subsection*{Funding}
Jihong Park's research was supported by the BK21 FOUR 
(Fostering Outstanding Universities for Research, No.~5120200913674), 
funded by the Ministry of Education (MOE, Korea) and the National Research 
Foundation of Korea (NRF).
Yire Shin's research was supported by the Basic Science Research Program through the National Research Foundation of Korea (NRF), funded by the Ministry of Education (RS-2025-25436608).

\subsection*{Conflict of interest}
The authors declare that they have no conflicts of interest.

	\subsection*{Use of generative artificial intelligence}	
Generative artificial intelligence tools, including ChatGPT (OpenAI) and
Claude (Anthropic),
were used to assist with theoretical development, proof
strategies, mathematical derivations, and manuscript preparation. The tools were not
used to generate data or numerical results.
All results, proofs, references, and conclusions were reviewed and
verified by the authors, who take full responsibility for the final
content.

\subsection*{Author contributions}
All authors contributed to the conception and design of the study, data 
collection, material preparation, and analysis. The first draft of the 
manuscript was written by Shin and J-S Park, and the other authors commented 
on previous versions of the manuscript. All authors read and approved the 
final manuscript.

\appendix
\section*{Appendix}

\section{Proof of Proposition~\ref{prop:spacing_ccdf}}

\begin{proof}
At the fixed true parameter value $\theta_0$, the standard
$r$-largest Gumbel representation implies that, for $r\ge2$, both
$C_r$ and $S_r$ follow the standard exponential distribution
\citep{Weissman1978,tawn1988extreme}. So $E(C_r)=E(S_r)=1$ and $\mathrm{Var}(C_r)=\mathrm{Var}(S_r)=1$.
	
	Let
	\[
	B_r=\frac{Z_{r-1}}{Z_r}.
	\]
	Since $Z_{r-1}\sim\mathrm{Gamma}(r-1,1)$ and
	$C_r=Z_r-Z_{r-1}\sim\mathrm{Gamma}(1,1)$ are independent, the
	beta--gamma relationship gives $B_r\sim\mathrm{Beta}(r-1,1)$ with $B_r\perp Z_r$. Also, $S_r=-(r-1)\log B_r$ and $C_r=Z_r(1-B_r)$.
	
	For $B_r\sim\mathrm{Beta}(r-1,1)$,
	\[
	E(B_r^a)=\frac{r-1}{r-1+a}.
	\]
	Differentiation with respect to $a$ gives
	\[
	E(-\log B_r)=\frac{1}{r-1},
	\qquad
	E(-B_r\log B_r)=\frac{r-1}{r^2}.
	\]
	Hence, using $E(Z_r)=r$,
	\[
	E(S_rC_r)
	=
	(r-1)r
	\left\{
	\frac{1}{r-1}-\frac{r-1}{r^2}
	\right\}
	=
	\frac{2r-1}{r}.
	\]
	It follows that
	\[
	\mathrm{Corr}(S_r,C_r)
	=
	E(S_rC_r)-1
	=
	1-\frac{1}{r},
	\]
	which proves Eq.~(\ref{eq:corr_sc}). Similarly,
	\[
	E\{(S_r-C_r)^2\}
	=
	2+2-2E(S_rC_r)
	=
	\frac{2}{r},
	\]
	proving Eq.~(\ref{eq:mse_sc}). Finally, Chebyshev's inequality yields
	\[
	P(|S_r-C_r|>\epsilon)
	\le
	\frac{2}{r\epsilon^2}.
	\]
	Combining this with the trivial upper bound of one proves
	Eq.~(\ref{eq:prob_sc}).
\end{proof}

\section{Transformation of the $r$GEV model to the $r$-largest
	standard Gumbel distribution}
\label{app:gumbel_transform}

Let $\underline{X}^{\,r}=(X^{(1)},\ldots,X^{(r)})$ follow the
$r$GEV distribution in Eq.~(\ref{rgevd}), and define
\[
Y_j
=
-\frac{1}{k}
\log\left\{
1-k\frac{X^{(j)}-\mu}{\sigma}
\right\},
\qquad j=1,\ldots,r,
\]
with the case $k=0$ understood by continuity.

For $k\ne0$, the inverse transformation is
\[
X^{(j)}
=
\mu+\frac{\sigma}{k}
\{1-\exp(-kY_j)\},
\]
so that
\[
w(X^{(j)})=\exp(-kY_j),
\qquad
\left|
\frac{\partial X^{(j)}}{\partial Y_j}
\right|
=
\sigma\exp(-kY_j).
\]
Since the transformation is componentwise, substitution into
Eq.~(\ref{rgevd}) gives
\[
f_{\boldsymbol Y}(y^{(1)},\ldots,y^{(r)})
=
\exp\{-\exp(-y^{(r)})\}
\prod_{j=1}^{r}\exp(-y^{(j)}),
\]
for $y^{(1)}>\cdots>y^{(r)}$. Thus, $(Y_1,\ldots,Y_r)$ has the $r$-largest
standard Gumbel distribution. It therefore follows from
Theorem~3 of \citet{Weissman1978} that the successive spacings $D_j=Y_{j-1}-Y_j, \quad j=2,\ldots,r$, are independent exponential random variables with mean
$1/(j-1)$. Consequently, $S_j=(j-1)D_j, \quad j=2,\ldots,r$, are independent standard exponential random variables.

 	\renewcommand{\baselinestretch}{1.0}
 
	\setlength{\bibsep}{1pt plus 0.3pt}
	\renewcommand{\baselinestretch}{1.0}\selectfont
	\bibliographystyle{spbasic}

	\bibliography{ref_rsel_CCDF_final}

@article{an2007r,
	author  = {An, Ying and Pandey, M. D.},
	year    = {2007},
	title   = {The r largest order statistics model for extreme wind speed estimation},
	journal = {J Wind Eng Ind Aerodyn},
	volume  = {95},
	number  = {3},
	pages   = {165--182},
	doi     = {10.1016/j.jweia.2006.05.008}
}

@article{Andrews2002,
	author  = {Andrews, Donald W. K.},
	title   = {Higher-Order Improvements of a Computationally Attractive $k$-Step Bootstrap for Extremum Estimators},
	journal = {Econometrica},
	year    = {2002},
	volume  = {70},
	number  = {1},
	pages   = {119--162},
	doi     = {10.1111/1468-0262.00271}
}

@misc{bader2020package,
	title={Package ‘eva’: Extreme Value Analysis with Goodness-of-Fit Testing, {V}ersion 0.2.6},
	author={Bader, Brian and Yan, Jun},
	url={https://github.com/brianbader/eva_package},
	year={2020}
}

@article{bader2017automated,
  title={Automated selection of $r$ for the $r$ largest order statistics approach with adjustment for sequential testing},
  author={Bader, Brian and Yan, Jun and Zhang, Xuebin},
  journal={Stat Comput},
  volume={27},
  number={6},
  pages={1435--1451},
  year={2017},
  publisher={Springer},
  doi={10.1007/s11222-016-9697-3}
}

@article{bader2018automated,
	title={Automated threshold selection for extreme value analysis via ordered goodness-of-fit tests with adjustment for false discovery rate},
	author={Bader, Brian and Yan, Jun and Zhang, Xuebin},
	journal={Ann Appl Stat},
	volume={12},
	number={1},
	pages={310--329},
	year={2018}
}

@book{beirlant2006book,
	title={Statistics of extremes: Theory and applications},
	author={Beirlant, J. and Goegebeur, Y. and Segers, J. and Teugels, J. L.},
	year={2006},
	publisher={John Wiley {and} Sons}
}

@techreport{BergBakken2007,
	author      = {Berg, Daniel and Bakken, Henrik},
	title       = {A copula goodness-of-fit approach based on the conditional probability integral transformation},
	institution = {Norwegian Computing Center},
	address     = {Norway},
	year        = {2007},
	url         = {www.danielberg.no/publications/Btest.pdf}
}

@article{BurkeGombay1988,
	author  = {Burke, Murray D. and Gombay, Edit},
	title   = {On Goodness-of-Fit and the Bootstrap},
	journal = {Statistics \& Probability Letters},
	year    = {1988},
	volume  = {6},
	number  = {5},
	pages   = {287--293},
	doi     = {10.1016/0167-7152(88)90001-6}
}

@article{Busaba2025CSAM,
	title={Hybrid estimation and entropy difference test in generalized logistic model for $r$-largest order statistics},
	author={Busababodhin, Piyapatr and Shin, Yire and Park, Jeong-Soo},
	journal={Commun Stat Appl Methods},
	volume={32},
	number={3},
	pages={325--339},
	year={2025}
}

@article{Busababodhin2026,
	author = {Piyapatr Busababodhin and Yire Shin and Jeong-Soo Park},
	title = {Hybrid estimation using {L}-moments for r-largest order statistics approach},
	journal = {Journal of Statistical Computation and Simulation, in press},
	year = {2026},
	publisher = {Taylor \& Francis},
	doi = {10.1080/00949655.2026.2736797}  
}

@article{BucherZhou2021,
	title={A Horse Race between the Block Maxima Method and the Peak–over–Threshold Approach},
	author={B{\"u}cher, Axel and Zhou, Chen},
	journal={Stat Sci},
	volume={36},
	number={3},
	pages={360--378},
	year={2021},
	doi={https://www.jstor.org/stable/27286449}
}

@book{CasellaBerger2nded,
	title={{Statistical inference, 2nd edn.}},
	author={Casella, George and Berger, Roger L.},
	year={2002},
	publisher={Duxbury},
    address={Pacific Grove}
}

@book{coles2001introduction,
  title={An introduction to statistical modeling of extreme values},
  author={Coles, Stuart},
  year={2001},
  publisher={Springer},
  address={London},
  doi={10.1007/978-1-4471-3675-0}
}

@article{DobricSchmid2007,
	author  = {Dobri{\'c}, Jadran and Schmid, Friedrich},
	title   = {A Goodness of Fit Test for Copulas Based on Rosenblatt's Transformation},
	journal = {Computational Statistics \& Data Analysis},
	year    = {2007},
	volume  = {51},
	pages   = {4633--4642},
	doi     = {10.1016/j.csda.2006.08.012}
}

@article{dupuis1997extreme,
  title={Extreme value theory based on the $r$ largest annual events: A robust approach},
  author={Dupuis, D J},
  journal={J Hydrol},
  volume={200},
  number={1-4},
  pages={295--306},
  year={1997},
  publisher={Elsevier},
  doi={10.1016/S0022-1694(97)00022-X}
}

@book{embrechts2013modelling,
	title={Modelling extremal events for insurance and finance},
	author={Embrechts, Paul and Kl{\"u}ppelberg, Claudia and Mikosch, Thomas},
	year={2013},
	publisher={Springer}
}

@article{FerroSegers2003,
	author  = {Ferro, Christopher A. T. and Segers, Johan},
	title   = {Inference for clusters of extreme values},
	journal = {Journal of the Royal Statistical Society: Series B (Statistical Methodology)},
	year    = {2003},
	volume  = {65},
	number  = {2},
	pages   = {545--556},
	doi     = {10.1111/1467-9868.00401}
}

@article{GenestRemillard2008,
	author  = {Genest, Christian and R{\'e}millard, Bruno},
	title   = {Validity of the Parametric Bootstrap for Goodness-of-Fit Testing in Semiparametric Models},
	journal = {Annales de l'Institut Henri Poincar{\'e}, Probabilit{\'e}s et Statistiques},
	year    = {2008},
	volume  = {44},
	number  = {6},
	pages   = {1096--1127},
	doi     = {10.1214/07-AIHP148}
}

@article{GenestQuessyRemillard2006,
	author  = {Genest, Christian and Quessy, Jean-Fran{\c{c}}ois
	and R{\'e}millard, Bruno},
	title   = {Goodness-of-Fit Procedures for Copula Models Based on the
	Probability Integral Transformation},
	journal = {Scandinavian Journal of Statistics},
	year    = {2006},
	volume  = {33},
	pages   = {337--366},
	doi     = {10.1111/j.1467-9469.2006.00470.x}
}

@article{gsell2016sequential,
	title={Sequential selection procedures and false discovery rate control},
	author={G'Sell, Max Grazier and Wager, Stefan and Chouldechova, Alexandra and Tibshirani, Robert},
	journal={J Royal Stat Soc B: Stat Method},
	volume={78},
	number={2},
	pages={423--444},
	year={2016},
	publisher={Oxford University Press}
}

@article{hamdi2014extreme,
  title={Extreme storm surges: A comparative study of frequency analysis approaches},
  author={Hamdi, Yasser and Bardet, L and Duluc, C-M and Rebour, V},
  journal={Nat Hazards Earth Syst Sci},
  volume={14},
  number={8},
  pages={2053--2067},
  year={2014},
  publisher={Copernicus GmbH},
  doi={10.5194/nhess-14-2053-2014}
}

@article{heinonen7451430dependence,
	title={On the dependence structure of the largest k order statistics},
	author={Heinonen, Helena and Ne\v{s}lehov\'a, Johanna},
	journal={Available at SSRN 7451430},
	year={2026}
}

@article{hofert2014graphical,
	title={A graphical goodness-of-fit test for dependence models in higher dimensions},
	author={Hofert, Marius and M{\"a}chler, Martin},
	journal={Journal of Computational and Graphical Statistics},
	volume={23},
	number={3},
	pages={700--716},
	year={2014},
	publisher={Taylor \& Francis}
}

@book{hosking1997regional,
  title={Regional frequency analysis: An
  approach based on L-moments},
  author={Hosking, Jonathan Richard Morley and Wallis, James R},
  year={1997},
  publisher={Cambridge University Press},
  address={Cambridge}
}

@article{houghton1978birth,
	title={Birth of a parent: The Wakeby distribution for modeling flood flows},
	author={Houghton, John C},
	journal={Water Resour Res},
	volume={14},
	number={6},
	pages={1105--1109},
	year={1978},
	publisher={Wiley Online Library}
}

@book{johnson1995continuous,
  title={{Continuous univariate distributions, Volume 2}},
  author={Johnson, Norman L and Kotz, Samuel and Balakrishnan, Narayanaswamy},
  year={1995},
  publisher={John Wiley \& Sons}
}

@article{kojadinovic2012goodness,
	title={Goodness-of-fit testing based on a weighted bootstrap: A fast large-sample alternative to the parametric bootstrap},
	author={Kojadinovic, Ivan and Yan, Jun},
	journal={Canadian J Statist},
	volume={40},
	number={3},
	pages={480--500},
	year={2012},
	publisher={Wiley Online Library}
}

@article{LandwehrMatalasWallis1979Wakeby,
	title={Estimation of parameters and quantiles of \text{W}akeby Distributions: 1. \text{K}nown lower bounds},
	author={Landwehr, J. M. and  Matalas, N. C. and Wallis, J. R. },
	journal={Water Resour Res},
	year={1979},
	volume = {15},
    number = {6},
	pages={1361--1372},
    doi={10.1029/WR015i006p01361}
}

@inproceedings{LeCam1956,
	author    = {Le Cam, Lucien},
	title     = {On the Asymptotic Theory of Estimation and Testing Hypotheses},
	booktitle = {Proceedings of the Third Berkeley Symposium on Mathematical Statistics and Probability},
	volume    = {1},
	pages     = {129--156},
	year      = {1956},
	publisher = {University of California Press},
	address   = {Berkeley, CA}
}

@article{muhammed2017variations,
  title={{Variations in return value estimate of ocean surface waves--A study based on measured buoy data and ERA-Interim reanalysis data}},
  author={Naseef, Muhammed T and  Kumar, Sanil V},
  journal={Nat Hazards Earth Syst Sci},
  volume={17},
  number={10},
  pages={1763--1778},
  year={2017},
  publisher={Copernicus GmbH}
}

@article{rosenblatt1952remarks,
	title={Remarks on a multivariate transformation},
	author={Rosenblatt, Murray},
	journal={Ann Math Statist},
	volume={23},
	number={3},
	pages={470--472},
	year={1952},
	publisher={JSTOR}
}

@article{silva2022dynamic,
  title={{Dynamic linear seasonal models applied to extreme temperature data: {A Bay}esian approach using the $r$-largest order statistics distribution}},
  author={Silva, Renato Santos da and Nascimento, Fernando Ferraz do and Bourguignon, Marcelo},
  journal={J Stat Comput Simul},
  volume={92},
  number={4},
  pages={705--723},
  year={2022},
  publisher={Taylor \& Francis}
}

@article{e2020change,
  title={A change-point model for the $r$-largest order statistics with applications to environmental and financial data},
  author={e Silva, Wyara Vanesa Moura and Nascimento, Fernando Ferraz do and Bourguignon, Marcelo},
  journal={Appl Math Model},
  volume={82},
  pages={666--679},
  year={2020},
  publisher={Elsevier}
}

@article{smith1986extreme,
  title={Extreme value theory based on the $r$ largest annual events},
  author={Smith, Richard L},
  journal={J Hydrol},
  volume={86},
  number={1-2},
  pages={27--43},
  year={1986},
  publisher={Elsevier}
}

@article{soares2004application,
  title={Application of the $r$ largest-order statistics for long-term predictions of significant wave height},
  author={Soares, C Guedes and Scotto, MG},
  journal={Coast Eng},
  volume={51},
  number={5-6},
  pages={387--394},
  year={2004},
  publisher={Elsevier}
}

@article{Stutetal1993,
	author  = {Stute, Winfried and Gonz{\'a}lez-Manteiga, Wenceslao
	and Presedo-Quindimil, Manuel Antonio},
	title   = {Bootstrap Based Goodness-of-Fit-Tests},
	journal = {Metrika},
	year    = {1993},
	volume  = {40},
	number  = {1},
	pages   = {243--256},
	doi     = {10.1007/BF02613687}
}

@article{Tabari_dilemma2021,
   title = {Extreme value analysis dilemma for climate change impact assessment on global flood and extreme precipitation},
   journal = {J Hydrol},
   volume = {593},
   eid = {125932},
   year = {2021},
   issn = {0022-1694},
   author = {Hossein Tabari}
}

@article{tawn1988extreme,
  title={An extreme-value theory model for dependent observations},
  author={Tawn, Jonathan A},
  journal={J Hydrol},
  volume={101},
  number={1-4},
  pages={227--250},
  year={1988},
  publisher={Elsevier}
}

@article{unnikrishnan2004analysis,
  title={Analysis of extreme sea level along the east coast of \text{I}ndia},
  author={Unnikrishnan, A S and Sundar, D and Blackman, D},
  journal={J Geophys Res: Oceans}, 
  volume={109},
  number={C6},
  pages={C0{60}23},
  year={2004},
  publisher={Wiley Online Library}
}

@book{vanderVaart1998,
	author    = {van der Vaart, Aad W.},
	title     = {Asymptotic Statistics},
	year      = {1998},
	publisher = {Cambridge University Press},
	address   = {Cambridge},
	doi       = {10.1017/CBO9780511802256}
}

@article{wang2008downscaling,
  title={{Downscaling and projection of winter extreme daily precipitation over North America}},
  author={Wang, Jiafeng and Zhang, Xuebin},
  journal={J Clim},
  volume={21},
  number={5},
  pages={923--937},
  year={2008},
  publisher={American Meteorological Society}
}

@article{wang1995selection,
	title={Selection of the k largest order statistics for the domain of attraction of the {Gumbel} distribution},
	author={Wang, Julian Z},
	journal={J Am Stat Assoc},
	volume={90},
	number={431},
	pages={1055--1061},
	year={1995},
	publisher={Taylor \& Francis}
}

@article{Weissman1978,
 author = {Ishay Weissman},
 journal = {J Am Stat Assoc},
 number = {364},
 pages = {812--815},
 publisher = {[American Statistical Association, Taylor & Francis, Ltd.]},
 title = {Estimation of Parameters and Larger Quantiles Based on the $k$ Largest Observations},
 volume = {73},
 year = {1978}
}

@article{Durbin1973,
  author  = {Durbin, James},
  title   = {Weak convergence of the sample distribution function when parameters are estimated},
  journal = {Ann Stat},
  volume  = {1},
  number  = {2},
  pages   = {279--290},
  year    = {1973}
}

@article{shin2023modeling,
  title={Modeling climate extremes using the four-parameter kappa distribution for $r$-largest order statistics},
  author={Shin, Yire and Park, Jeong-Soo},
  journal={Weather Clim Extrem},
  volume={39},
  eid={10{05}33},
  year={2023},
  publisher={Elsevier},
  note={{R}evision at arXiv.2007.12031}
}

@article{ShinPark2024Serra,
  title={Generalized logistic model for $r$ largest order statistics, with hydrological application},
  author={Shin, Yire and Park, Jeong-Soo},
  journal={Stoch Environ Res Risk Assess},
  volume={38},
  number={4},
  pages={1567--1581},
  year={2024}
}

@misc{ismev2025,
  author = {Heffernan, Janet E. and Gilleland, Eric},
  title  = {ismev: An Introduction to Statistical Modeling of Extreme Values},
  year   = {2025},
  note   = {R package version 1.43, original S functions by S. Coles. DOI 10.32614/CRAN.package.ismev}
}

@misc{ShinParkRpackageevmr,
	title={evmr: An {R} package for Extreme Value Modeling for r-Largest Order Statistics, Version 0.2.0},
	author={Shin, Yire and Park, Jeong-Soo},
	year={2026},
	url={https://cran.rstudio.com}
}

@article{Smith1985,
	author  = {Smith, Richard L.},
	title   = {Maximum Likelihood Estimation in a Class of Nonregular Cases},
	journal = {Biometrika},
	year    = {1985},
	volume  = {72},
	number  = {1},
	pages   = {67--90},
	doi     = {10.1093/biomet/72.1.67}
}

@article{DavidsonMacKinnon1999,
	author  = {Davidson, Russell and MacKinnon, James G.},
	title   = {Bootstrap testing in nonlinear models},
	journal = {International Economic Review},
	year    = {1999},
	volume  = {40},
	number  = {2},
	pages   = {487--508}
}

@article{SalibianBarreraZamar2002,
	author  = {Salibi{\'a}n-Barrera, Mat{\'i}as and Zamar, Ruben H.},
	title   = {Bootstrapping robust estimates of regression},
	journal = {Ann Stat},
	year    = {2002},
	volume  = {30},
	number  = {2},
	pages   = {556--582}
}

@article{csorgHo1996exact,
	author  = {Cs{\"o}rg{\H{o}}, S{\'a}ndor and Faraway, Julian J.},
	title   = {The exact and asymptotic distributions of {C}ram{\'e}r--von {M}ises statistics},
	journal = {J R Stat Soc Ser B Methodol},
	year    = {1996},
	volume  = {58},
	number  = {1},
	pages   = {221--234}
}
	
\end{small}

\clearpage
\includepdf[pages=-]{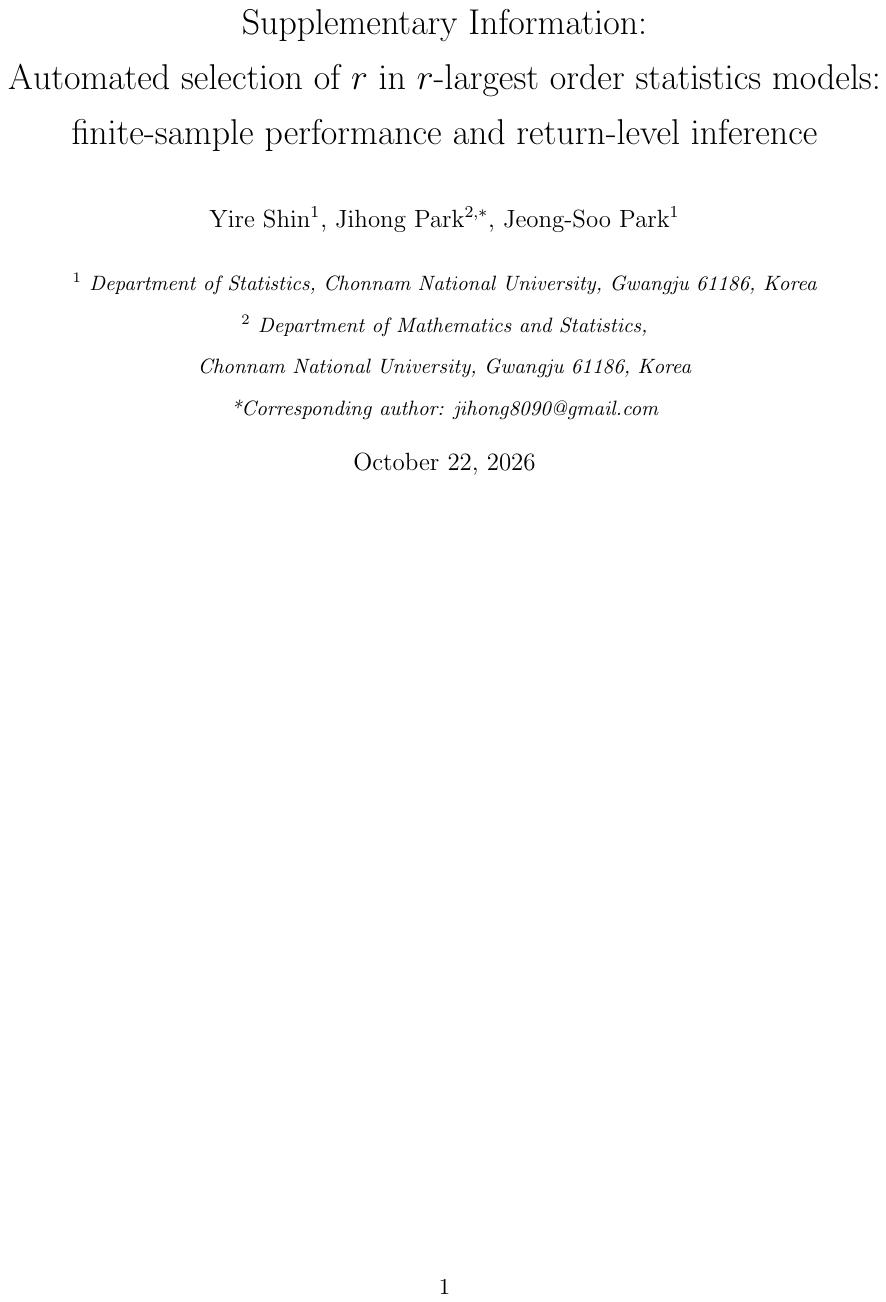}

\end{document}